\documentclass{aa}  
\newcommand{\micron}{$\mu$m}

\newcommand{\density}{cm$^{-3}$}
\newcommand{\gasdensity}{g cm$^{-3}$}
\newcommand{\water}{H$_2$O}
\newcommand{\methanol}{CH$_3$OH}

\newcommand{\spitzer}{{\it Spitzer}}
\newcommand{\herschel}{{\it Herschel}}

\newcommand{\gcm}{g cm$^{-2}$}

\newcommand{\radex}{\textsc{radex}}

\usepackage{graphicx}
\usepackage{txfonts}

\begin{document} 

   \title{ H$_2$O distribution in the disc of HD 100546 and HD 163296:\\  
    the role of dust dynamics and planet--disc interaction}    

   \author{
   L.M.\ Pirovano\inst{\ref{inst_unimi}},
        D.\ Fedele\inst{\ref{inst_inaf}},
    E.F. \ van Dishoeck\inst{\ref{inst_leiden},\ref{inst_mpe}},
        M.R. \ Hogerheijde\inst{\ref{inst_leiden},\ref{inst_amsterdam}},
        G. Lodato\inst{\ref{inst_unimi}},
    S. \ Bruderer\inst{\ref{inst_mpe}}
}

\institute{
Dipartimento di Fisica, Universit\'a degli Studi di Milano, Via Giovanni 
 Celoria, 16, I-20133 Milano, MI, Italia\label{inst_unimi}
\and
INAF-Osservatorio Astrofisico di Arcetri, L.go E. Fermi 5, I-50125 Firenze, Italy\label{inst_inaf}
\and
Leiden Observatory, Leiden University, P.O. Box 9513, NL-2300 RA, Leiden, The Netherlands\label{inst_leiden}
\and
Anton Pannekoek Institute for Astronomy, University of Amsterdam, Science Park 904, 1098 XH, Amsterdam, The Netherlands\label{inst_amsterdam}
\and
Max Planck Institut f\"{u}r Extraterrestrische Physik, Giessenbachstrasse 1, 85748 Garching, Germany\label{inst_mpe}
\\
}
\titlerunning{Water in HD 100546 and HD 163296}      
\authorrunning{Pirovano et al.}

\abstract{
Water plays a fundamental role in the formation of planets and their atmospheres. Far-infrared observations with the {\it Herschel} Space Observatory revealed a surprisingly low abundance of cold-water reservoirs in protoplanetary discs. On the other hand, a handful of discs show emission of hot water transitions excited at temperatures above a few hundred Kelvin. In particular, the protoplanetary discs around the Herbig Ae stars HD 100546 and HD 163296 show opposite trends in terms of cold versus hot water emission: in the first case, the ground-state transitions are detected and the high-$J$ lines are undetected, while the trend is opposite in HD 163296. As the different transitions arise from different regions of the disc, it is possible to address the overall distribution of water molecules throughout the disc. We performed a detailed spectral analysis using the thermo-chemical model DALI. We find that HD 163296 is characterised by a water-rich (abundance $\gtrsim 10^{-5}$) hot inner disc (within the snowline) and a water-poor ($< 10^{-10}$) outer disc: the relative abundance of water molecules in the hot inner region may be due to the thermal desorption of icy grains that have migrated inward. Remarkably, the size of the \water \ emitting region corresponds to a narrow dust gap visible in the millimeter continuum at $r=10\,$au observed with the Atacama Large Millimetre Array (ALMA). This spatial coincidence may be due to pebble growth at the border of the snow line. The low-$J$ lines detected in HD 100546  instead imply an abundance of a few $10^{-9}$ in the cold outer disc ($> 40\,$au). The emitting region of the cold H$_2$O transitions is spatially coincident with that of the H$_2$O ice previously seen in the near-infrared. Notably, millimetre observations with ALMA reveal the presence of a large dust gap between nearly 40 and 150\,au, likely opened by a massive embedded protoplanet.   
In both discs, we find that the warm molecular layer in the outer region (beyond the snow line) is highly depleted of water molecules, implying an oxygen-poor chemical composition of the gas. We speculate that gas-phase oxygen in the outer disc is readily depleted and its distribution in the disc is tightly coupled to the dynamics of the dust grains.  
}



 \maketitle

\section{Introduction}
Water is a key element in the physical and chemical evolution of protoplanetary discs and in the formation of planets \citep[e.g. ][]{vanDishoeck21}. 
Ro-vibrational and rotational transitions of \water \ and OH in discs have been found by several authors using \spitzer \ 
\citep{Carr08, Salyk08, Pontoppidan10a}, Keck/NIRSPEC \citep{Salyk08, Mandell08, Doppmann11}, VLT/CRIRES 
\citep{Fedele11, Banzatti17}, VLT/VISIR \citep{Pontoppidan10a}, and \herschel \ \citep{Hogerheijde11, Riviere12, Meeus12, Fedele12, Fedele13a, Podio13}.
The detection of \water \ and OH triggered intensive modelling campaigns designed to improve our understanding of the origin and abundance of water in
discs. Various chemical and physical--chemical models have been developed, including different processes: \citet{Glassgold09} and \citet{Najita11}
focus on the formation of \water \ in the inner disc and take into account the effect of the energetic radiation 
(Ultraviolet and X-rays); \citet{Bethell09} and \citet{Adamkovics14} included molecular self-shielding to explain the high abundance of \water \ in
the inner region of T Tauri discs and the physical--chemical models of \citet{Woitke09, Bruderer12, Du14} include gas-phase \water \
formation in the entire disc and in the disc atmosphere including desorption from icy grains.

Observations of multiple transitions of \water \ in protoplanetary discs have the potential to unveil the
abundance and distribution of water vapour ---both warm and cold--- in discs at the time of planet formation, as shown by \citet{Zhang13} for the disc around TW Hya. These authors find a high concentration of water vapour at $r \sim 4\,$au, corresponding to the water condensation front (or {\it snow line}).
This latter experiment was possible thanks to the combination of low- and high-$J$ \water \ transitions from \herschel \ and \spitzer.

This paper presents a \water \ `line mapping' of the two Herbig Ae systems where rotational \water \ emission has been detected with \herschel, namely HD 100546 and HD 163296. 
Both discs show substructures (cavities and dust rings) in the dust continuum with ALMA, which may or may not trap icy pebbles in the outer disc that could otherwise drift inward.
Our analysis is based on a mixture of \water \ lines targeted with the Photodetector Array Camera and Spectrometer (PACS) \citep[][]{Poglitsch10} and the Heterodyne Instrument for the Far Infrared (HIFI)  \citep[][]{deGraauw10} on \herschel.
The aim of this paper is to determine the abundance distribution of \water \ vapour in HD 100546 and HD 163296 by comparing the observed \water \ spectrum to a set of physical--chemical models of discs.


\begin{table}[t]\centering \caption{Observations log. \label{obs_id}}
\begin{tabular}{llll}
\hline
\hline
Source & Date & Obs ID & PI\\
\hline
\multicolumn{4}{c}{\bf HIFI - $1_{10}-1_{01}$}\\

 HD 163296 & 2012-10-12 & 1342253130 & L. Podio \\
 HD 100546 & 2011-12-23 & 1342235094 & M. Hogerheijde \\
 HD 100546 & 2011-12-24 & 1342235095 & M. Hogerheijde\\

\multicolumn{4}{c}{\bf HIFI - $1_{11}-0_{00}$}\\

 HD 163296 & 2012-10-17 & 1342253594 & M. Hogerheijde\\
 HD 163296 & 2012-10-18 & 1342253595 & M. Hogerheijde\\
 HD 100546 & 2012-07-24 & 1342248515 & M. Hogerheijde\\
 HD 100546 & 2012-12-04 & 1342256430 & M. Hogerheijde\\
 HD 100546 & 2012-12-27 & 1342257850 & M. Hogerheijde\\

\multicolumn{4}{c}{\bf PACS - $3_{21}-2_{12}$}\\

 HD 100546 & 2013-02-10 & 1342263460 & Calibration\\

\multicolumn{4}{c}{\bf PACS - Others}\\

 HD 163296 & 2011-04-03 & 1342217819 & N. Evans\\
 HD 163296 & 2011-04-03 & 1342217820 & N. Evans \\
 HD 100546 & 2009-12-11 & 1342188037 & N. Evans \\
 HD 100546 & 2009-12-11 & 1342188038 & N. Evans \\
\hline
\hline
\end{tabular}\label{tab:log}
\end{table}

\section{Water reservoirs in discs}\label{sec:waterintro}
According to physical--chemical models \citep[e.g.][]{Woitke09, Bruderer12, Du14}, there are three main reservoirs 
of gaseous H$_2$O in discs:

\begin{enumerate}

\item[$s_1$] is the inner reservoir (inside the snow line) where \water \ molecules are formed through gas-phase reactions. The high
kinetic temperature in this region allows the energy barrier of the radical-H$_2$ reactions to be overcome \citep[e.g.][]{Glassgold09}. 
At the same time, this reservoir is fully shielded from the ultraviolet (UV) radiation coming from the star, meaning that \water \ molecules 
are protected from photodissociation. The \water \ abundance in this region can be further enhanced if icy planetesimals migrate 
inwards through the snow line and \water \ molecules are released via thermal desorption.\\

\item[$s_2$] is the (cold) photodesorption layer in the outer disc where the \water \ vapour is released from the icy grains. The gas temperature in this region is too low ($< 100\,$K) and radical H$_2$ reactions are inhibited, meaning that the \water \ vapour reservoir is only supplied by photodesorption.\\

\item[$s_3$] is the warm reservoir which is located at intermediate layers above the disc midplane and that can extend to several astronomical units (au) from 
the star. The production of \water \ molecules is driven by gas-phase reactions. UV photons are not fully absorbed and photodissociation
of \water \ molecules is relevant in this region: because of this, the amount of \water \ vapour in this layer is also regulated by the
self-shielding \citep[e.g.][]{Bethell09}.\\

\end{enumerate}

The three \water-rich regions are described in Sect.~\ref{sec:dali}.
Each region contributes to the \water \ line flux, but while the high-$J$ rotational lines and the ro-vibrational lines arise primarily from reservoirs $s_1$ and $s_3$,  the low-$J$ rotational transitions and in particular those connecting to the ground state, are more sensitive to the cold photodesorption layer $s_2$.

\section{Observations}\label{sec:obs}
\subsection{PACS spectra}
HD 163296 and HD 100546 were observed with \herschel/PACS as part of the DIGIT \citep{Green13} and GASPS \citep{Dent13} key programs. 
The PACS spectra used here are from the DIGIT program and the observations are presented in \citet{Fedele12, Fedele13a}. The spectral window of \herschel/PACS (50-200\,\micron) includes several rotational \water \ transitions ranging in upper level energy from $E_{\rm up} = 114\,$K ($2_{12}-1_{01}$ at 179.52\,\micron) to $E_{\rm up}=1552\,$K ($9_{18}-9_{09}$ at 62.93\,\micron). The fluxes and upper limits of the \water \ transitions are reported in \citet{Fedele12, Fedele13a}. A number of transitions are detected towards HD 163296 \citep{Fedele12, Meeus12}, while no \water \ emission is detected in the DIGIT and GASPS programs towards HD 100546 \citep[][]{Fedele13a}. The PACS spectrum of HD 100546 was taken early in the mission \citep[][]{Sturm10} and has a lower signal-to-noise ratio (S/N) compared to that of HD 163296. The upper limits reported by \citet{Fedele13a} are almost an order of magnitude higher than the line fluxes measured for HD 163296.



\begin{figure}[t]
\centering
\includegraphics[width=9cm]{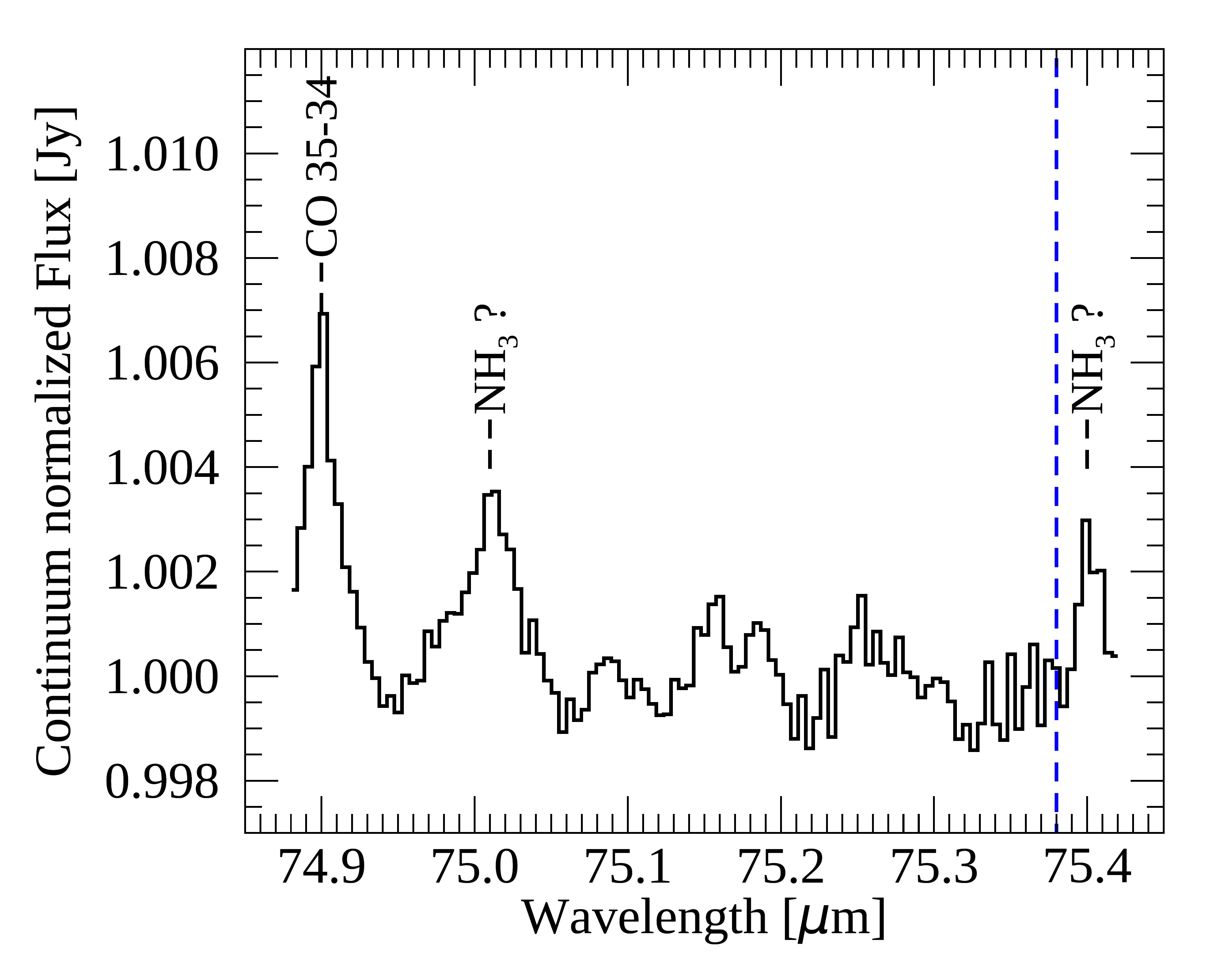}
\caption{{\it Herschel}/PACS deep integration of HD 100546 at 75\,\micron . The CO $J=35-34$ line is detected along with two other emission lines likely due to NH$_3$. The blue dashed line indicates the position of the \water \ ${3_{21} - 2_{12}}$ transition.}\label{fig:pacs}
\end{figure}

\smallskip
\noindent
A deep PACS observation of HD 100546 was executed as part of a calibration program targeting the \water \ transition $3_{21}-2_{12}$ at 75.38\,\micron . Figure \ref{fig:pacs} shows the pipeline-reduced spectrum (taken in `linespec' mode). 
The CO $J=35-34$ transition at 74.9 \micron \ is detected along with two other emission lines at 75.01 \micron \ and 75.41 \micron. These two emission lines could be due to ro-vibrational transitions of NH$_3$ or SO$_2$ (further analysis is deferred to a forthcoming paper). Notably, the \water \ $3_{21}-2_{12}$ is not detected. 
The 3$\sigma$ upper limit (Table \ref{tab:obs}) is computed adopting a line FWHM of 0.019~\micron\ as measured on the CO $J=35-34$ transition and accounting for the flux calibration error provided by the {\it Herschel} pipeline. 

  \begin{table}
        \caption{\water \ line fluxes.}
        \centering
        \begin{tabular}{lcccc}
        \hline\hline
        Transition     & $\lambda$ [\micron] & $E_{\rm up}$ [K] & \multicolumn{2}{c}{$\mathrm{F_{obs}}$ [$10^{-17}$ W\,m$^{-2}$]} \\
                       &                     &              &  HD 163296 & HD 100546   \\
        \hline
        $1_{10}-1_{01}$ & 538.30 & 61   & < 0.08        & 0.10 $\pm$ 0.01 \\
        $1_{11}-0_{00}$ & 269.54 & 53   & < 0.49        & 0.30 $\pm$ 0.02 \\
        $2_{21}-1_{10}$ & 108.07 & 194  & 0.7 $\pm$ 0.5 & < 10.2 \\
        $3_{21}-2_{12}$ & 75.38 &  305  & < 2.5         & < 1.09 \\
    $4_{14}-3_{03}$     & 113.54 & 324  & 0.7 $\pm$ 0.4 & < 7.7 \\
    $4_{23}-3_{12}$     & 78.74 & 432   & 1.8 $\pm$ 0.4 & < 14.7 \\
    $4_{31}-3_{22}$     & 56.31 &  552  & 2.7 $\pm$ 1.6 & < 23.6 \\
    $7_{07}-6_{16}$     & 71.95 & 843   & 2.2 $\pm$ 0.5 & < 11.8 \\
    $8_{18}-7_{07}$     & 63.32 & 1070  & 2.0 $\pm$ 0.6 & -- \\
        \hline
        \hline
        \end{tabular}\label{tab:obs}
        \tablefoot{The \water \ $8_{18}-7_{07}$ transition in HD 100546 is blended with the [\ion{O}{i}] line}
        \end{table}

\begin{figure}[t]
\centering
\includegraphics[width=9cm]{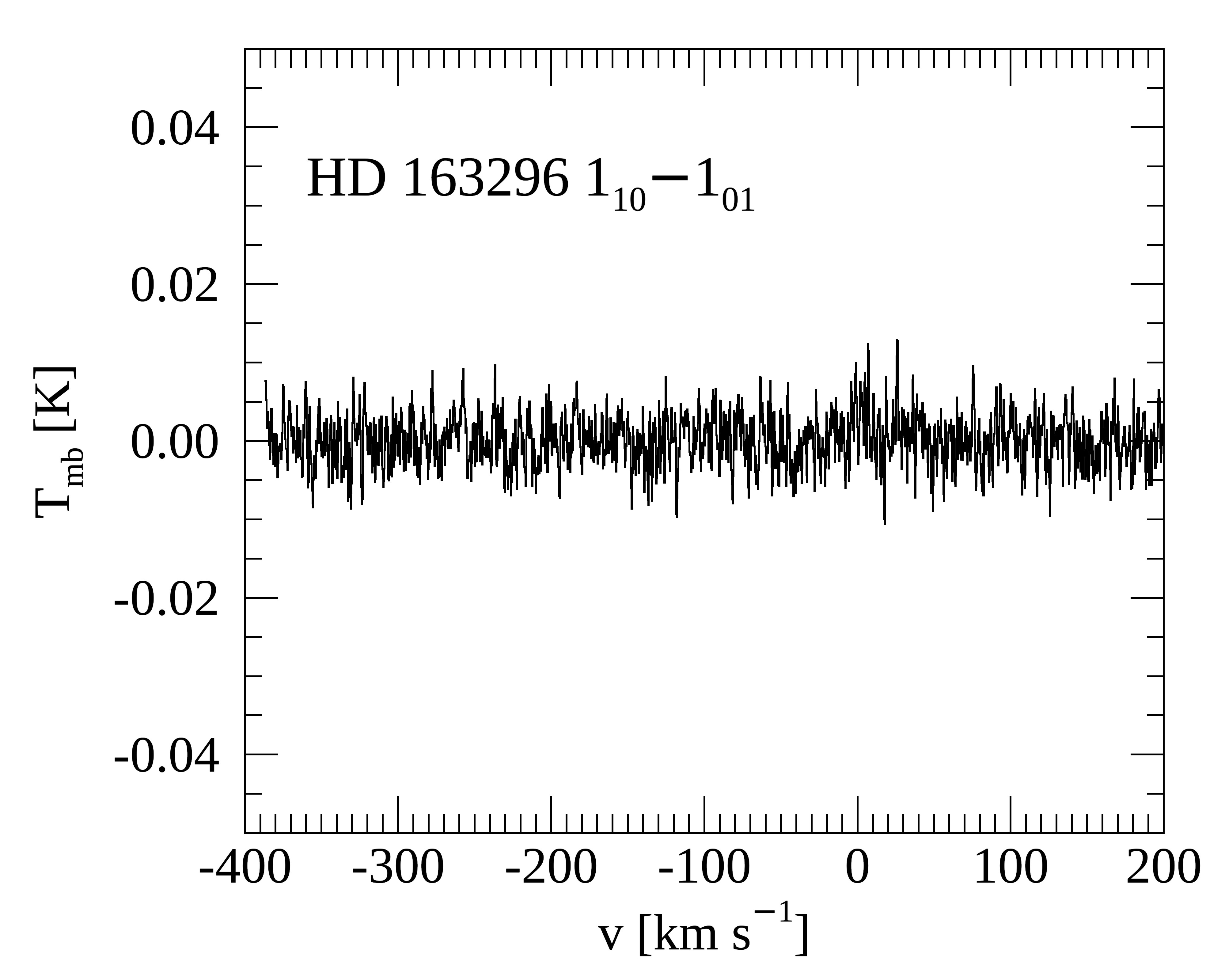}
\caption{{\it Herschel}/HIFI spectrum of HD 163296 $1_{10} - 1_{01}$ (at 556\,GHz) at rest-frame velocity.} \label{fig:hifi}
\end{figure}

\subsection{HIFI spectra}
The \water \ ground state transitions were observed with \herschel/HIFI during an open call proposal (Table~\ref{tab:log}). The line fluxes and upper limits were published by \citet{Du17}. Here we report a new, more stringent upper limit to the  $1_{10}-1_{01}$ transition towards HD 163296 based on deeper integration (Table \ref{tab:log}). We downloaded the level 2 data product from the data archive. The WBS spectrum is converted from antenna temperature to mean-beam temperature after correcting for the beam and forward efficiency \citep[][]{Roelfsema12}. Finally, the spectra of the horizontal and vertical polarization are averaged together. The reduced and flux-calibrated spectrum is presented in Figure \ref{fig:hifi}. The ground-state \water \ line remains undetected even in this case. We compute the upper limit and converted from K km~s$^{-1}$ to W~m$^{-2}$ using the formula 

\begin{equation}
\int F_{\nu} d\nu  = 2.73 \times 10^{-29} \nu^3 \theta^2 \int T_{\nu} dv 
,\end{equation}

\noindent where $\nu$ is the line frequency (GHz) and $\theta$ the half power beam width (arcsec). Following \citet{Du17}, we adopted a line full width at half maximum of 9~km~s$^{-1}$ \citep[e.g.][]{Thi04}, which is similar to the low-$J$ CO lines, and the line upper limits include a 20$\%$ flux calibration uncertainty.

  \begin{table*}
        \caption{DALI model parameters.}
        \centering
        \begin{tabular}{lcccl}
        \hline\hline
          Parameter             & Unit                & HD 163296 & HD 100546           &    Description \\
        \hline
        $M_{\star}$    & [$\mathrm{M}_{\odot}$]   & 2.0 & 2.5  & Stellar mass             \\
        $T_{\rm eff}$  & [K]             & 9000 & 10500        & Stellar (blackbody) temperature  \\
        $L_{\rm bol}$  & [$\mathrm{L}_{\odot}$]   & 17 & 28     & Stellar luminosity         \\
        $\Sigma_{\rm c}$ & [g cm$^{-2}$] & 1.55 & 18 & $\Sigma_{\rm gas}$ at $R_{\rm c}$ \\ 
        $\Delta_{\rm gd}$ & & 30 & 90 & Gas-to-dust mass ratio \\ 
        $\psi$         &                 & 0.10 & 0.23                  & Flaring angle       \\
        $h_{\rm c}$    & [radians]       & 0.07 & 0.18      & Scale height at $R_{\rm c}$         \\
        $\gamma$       &                 & 0.9 & 1                 & Power-law exponent of $\Sigma(r)$          \\
        $R_{\rm c}$    & [au]            & 125 & 50          & Critical radius  \\
        $R_{\rm in}$  & [au]              & 0.35 & 0.25     & Dust inner radius                  \\
        $R_{\rm out}$  & [au]            & 600    & 400    & Disc outer radius                \\
        $M_{\rm gas}$ &  [$10^{-2} \times \mathrm{M}_{\odot}$]      & 1.54 &  1.57 & Disc gas mass \\
        $M_{\rm dust}$ & [$10^{-4} \times \mathrm{M}_{\odot}$]   & 5.15 & 1.68  & Disc dust mass\\
        $\chi$         &                 & \multicolumn{2}{c}{0.2 }   & Degree of settling      \\
        $f_{\rm large}$&                 & \multicolumn{2}{c}{0.85} & Large-to-small dust mass ratio \\
        $a_{\rm min}$ & [\micron] & 0.02 & 0.05 & Minimum grain size\\
        $q$                 & & 3.0 & 3.5 & Power-law exponent of grain size population\\
        PAHs                & [\%] & 0.01 & 1.0 &  PAH abundance in gas with respect to the ISM                              \\     
        $R_{\rm cav, in} - R_{\rm cav, out}$  & [au] & & 4.0 - 13.0 & Dust cavity                  \\
        $\delta_{\rm gas, cav}$  &     &          & $10^{-5}$    & Gas density drop for $R_{\rm cav, in} < R < R_{\rm cav, out}$ \\
        $\delta_{\rm small, cav}$  &         &      & $10^{-5}$  & Small dust density drop for $R_{\rm cav, in} < R < R_{\rm cav, out}$ \\
        $\delta_{\rm large, cav}$  &       &        & $0$    & Large dust density drop for $R_{\rm cav, in} < R < R_{\rm cav, out}$ \\
        $R_{\rm gap, in} - R_{\rm gap, out}$  & [au]           & & 50 - 150   & Dust gap                  \\
        $\delta_{\rm gas, gap}$  &         &      & $10^{-1}$  & Gas density drop for $R_{\rm gap, in} < R < R_{\rm gap, out}$ \\
        $\delta_{\rm small, cav}$  &       &        & $10^{-1}$   & Small dust density drop for $R_{\rm gap, in} < R < R_{\rm gap, out}$ \\
        $\delta_{\rm large, cav}$  &        &       & $0$   & Large dust density drop for $R_{\rm gap, in} < R < R_{\rm gap, out}$ \\
        \hline
        \hline
        \end{tabular}\label{tab:dali}
        \end{table*}
        
\begin{figure*}
\centering
\includegraphics[width=18cm]{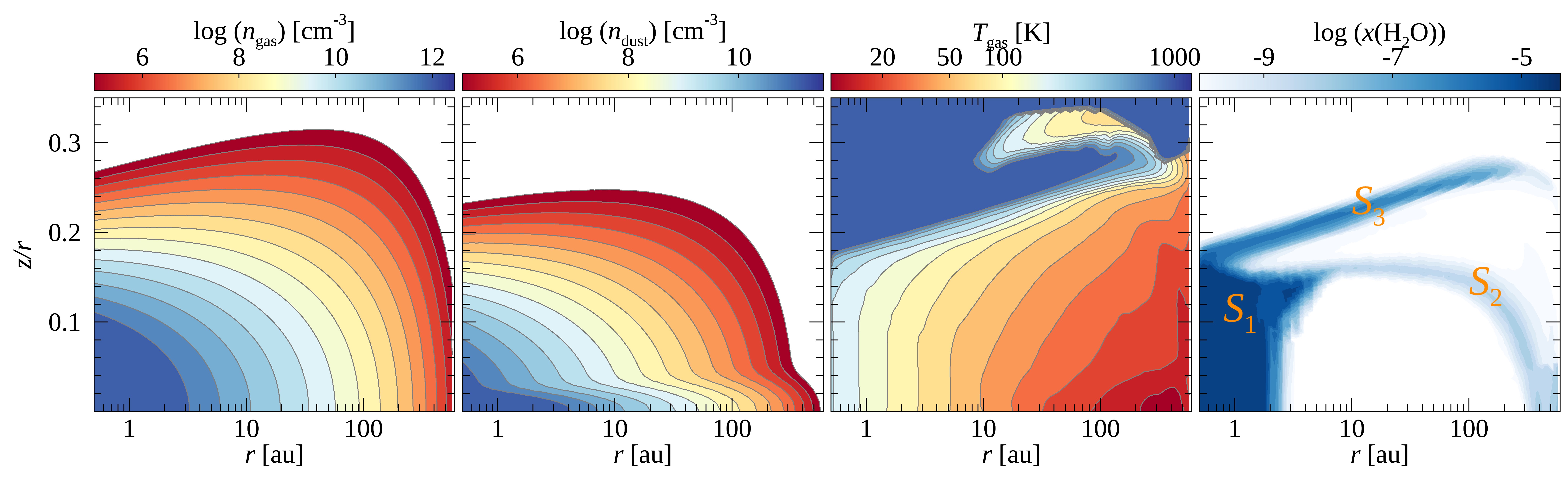}
\includegraphics[width=6cm]{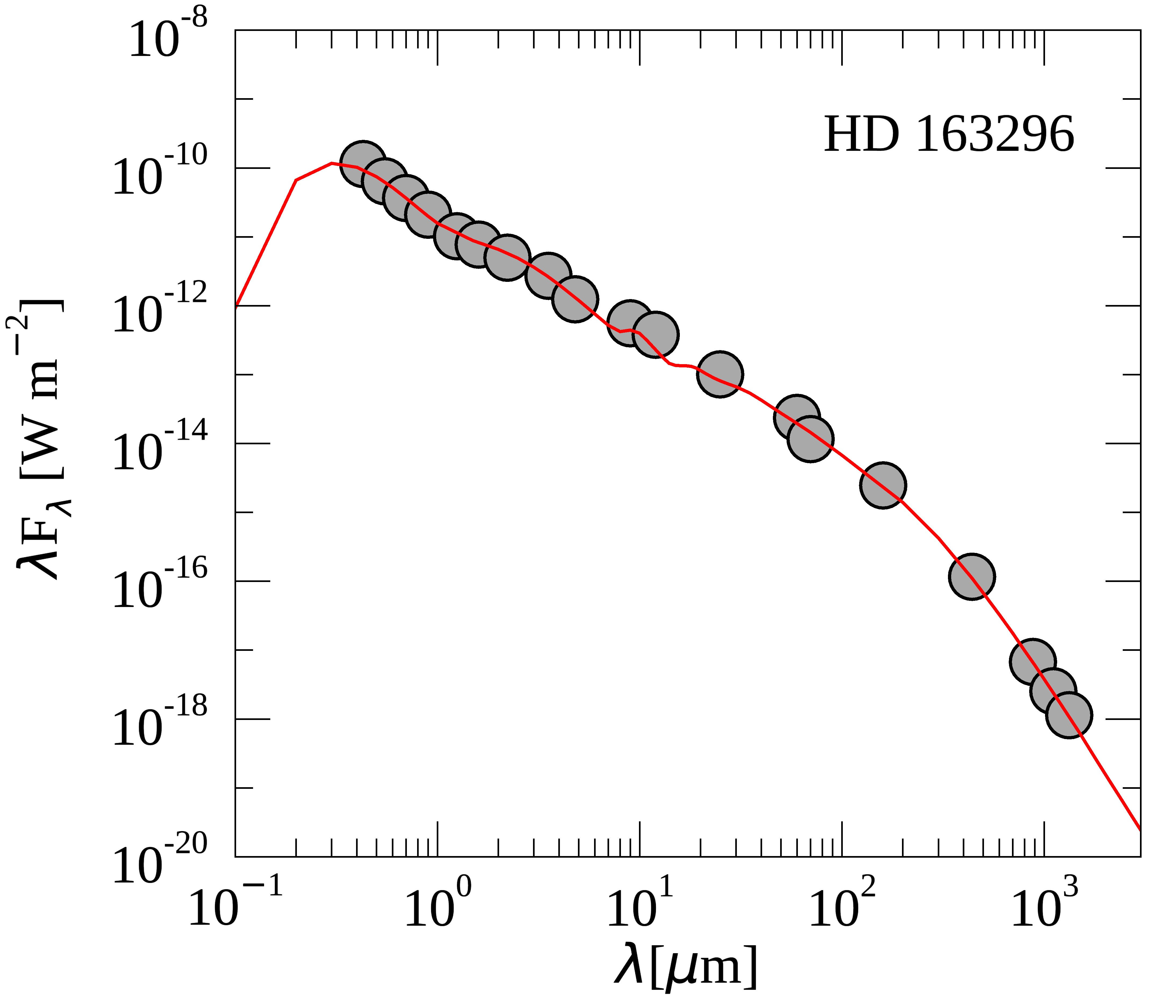}
\includegraphics[width=6cm]{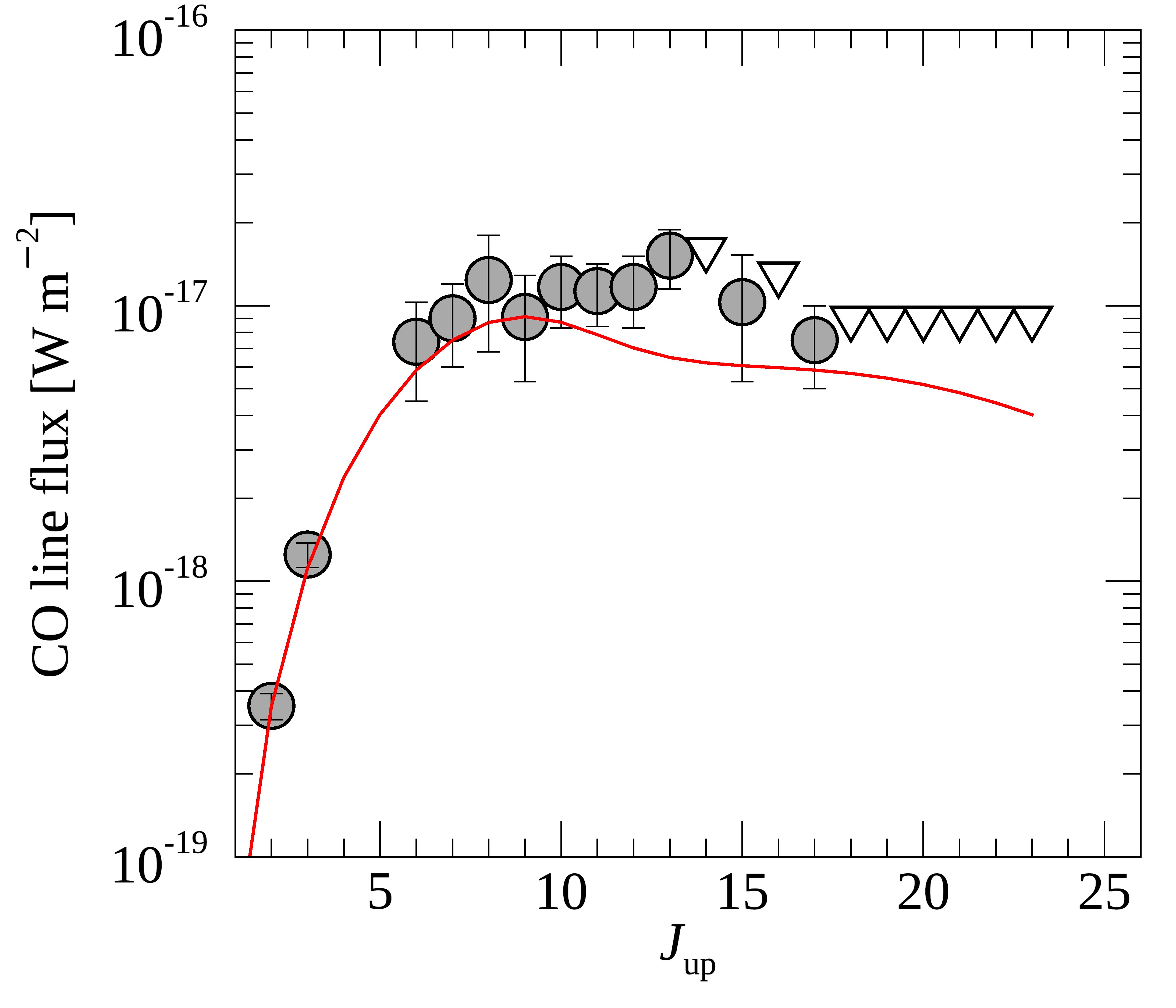}
\includegraphics[width=6cm]{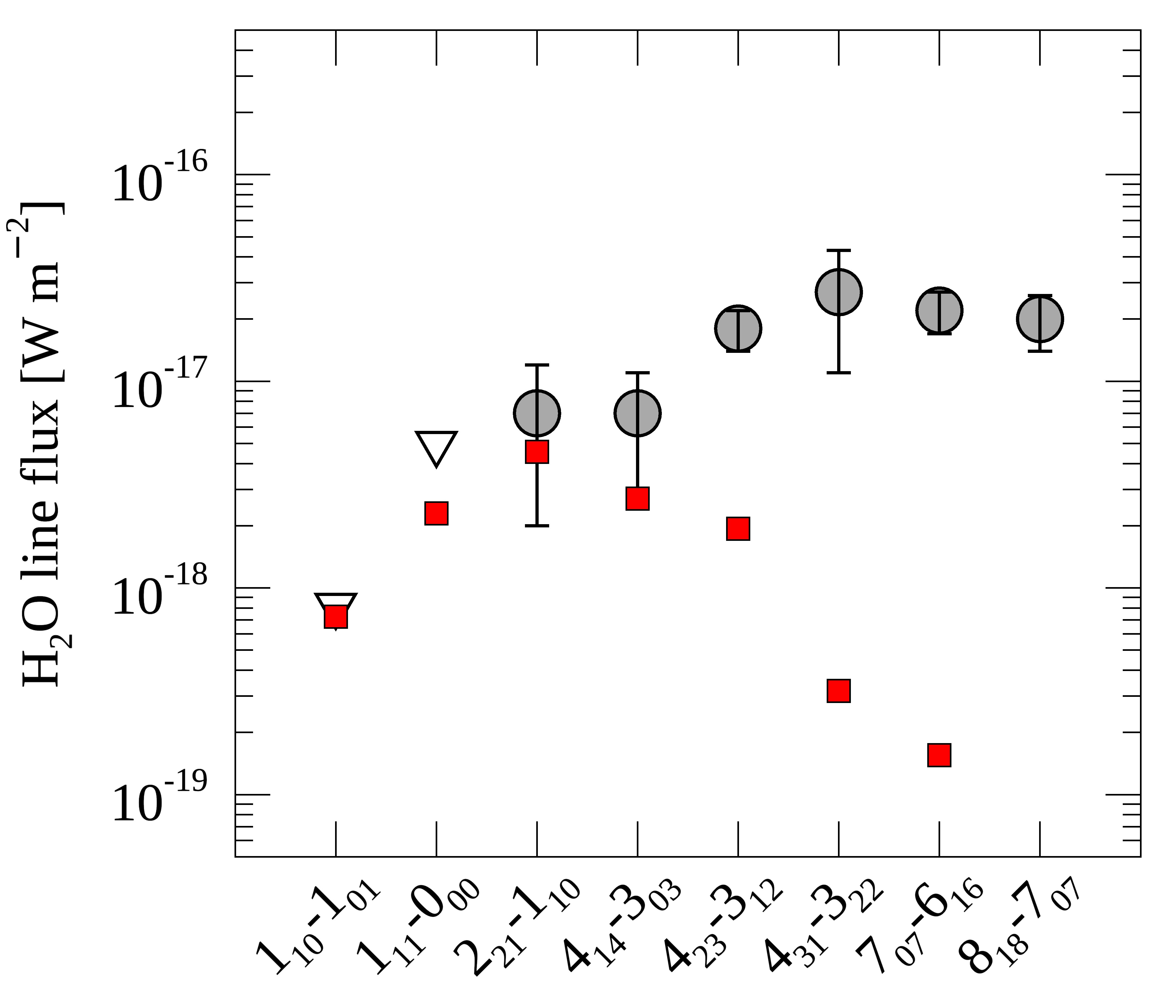}
\caption{DALI disc structure for HD 163296. The top panels  show the gas and dust density structure, the gas temperature, and the \water \ abundance structure. The three main \water \ reservoirs are indicated by the labels $s_1$, $s_2$, and $s_3$. The bottom panels show the model predictions of the SED (left), CO ladder (middle), and \water \ fluxes (right). The observed fluxes and 3$\sigma$ upper limits are shown as (grey) circles and open triangles, respectively. The representative model shown as a red line and red squares) reproduces the SED (left) and the CO rotational ladder (middle) well but is not able to reproduce the \water \ line fluxes and upper limits. We note that the \water \ abundances shown here are those of the representative disc model based on the full gas-grain chemistry calculation.}\label{fig:dali1}
\end{figure*}

\section{Analysis}\label{sec:dali}
The analysis presented in this paper is based on simulations with the DALI thermo-chemical models of discs \citep{Bruderer12, Bruderer13}. 

\subsection{Disc physical setup}
DALI takes as input the stellar spectrum (in this case corresponding to the blackbody emission at $T=T_{\rm eff}$) and a power-law surface density structure with an exponential tail:

\begin{equation}
\Sigma_{\rm gas}(R) = \Sigma_{\rm c}  \ \Bigg(\frac{R}{R_{\rm c}}\Bigg)^{-\gamma} \ \exp\Bigg[ - \Bigg( \frac{R}{R_{\rm c}} \Bigg)^{2 - \gamma} \Bigg]
\label{sigma_gas}
,\end{equation}

\noindent
where $R$ is the radial distance from the star, $R_{\rm c}$ the critical radius, and $\Sigma_{\rm c}$ is the gas surface density at $R=R_{\rm c}$. The dust surface density is $\Sigma_{\rm gas} / \Delta_{\rm gd}$, where $\Delta_{\rm gd}$ is the gas-to-dust-mass ratio. The density in the vertical direction is assumed to follow a Gaussian distribution with scale height $h = H/R:$

\begin{equation}
h = h_{\rm c} \Bigg( \frac{r}{R_{\rm c}}\Bigg)^{\psi} 
,\end{equation}

\noindent
where $h_{\rm c}$ is the scale height at $R=R_{\rm c}$ and $\psi$ the flaring angle. The dust grain size
distribution is described using two grain size populations as in \citet{dAlessio06}, and dust mass absorption
cross sections from \citet{Andrews11}. The grain size distribution is approximated by a power-law
distribution with exponent $q$ (equal for both  populations). The settling of the large grains is controlled 
by the parameters $\chi$ (scale height of the large grain population compared to the gas one) and $f_{\rm large}$ (small-to-large grain mass ratio).

DALI solves the 2D dust continuum radiative transfer and determines the dust temperature and radiation field strength at each disc position. In the standard procedure, DALI iteratively solves the gas thermal balance and chemistry. The \water \ spectrum is obtained by computing the non-local thermodynamic equilibrium (NLTE) level populations, including infrared pumping and by fixing the ortho-to-para ratio to 3. The collisional rate coefficients are from the LAMDA database \citep[][and references therein]{Schoier05}.

\begin{figure*}[t]
\centering
\includegraphics[width=18cm]{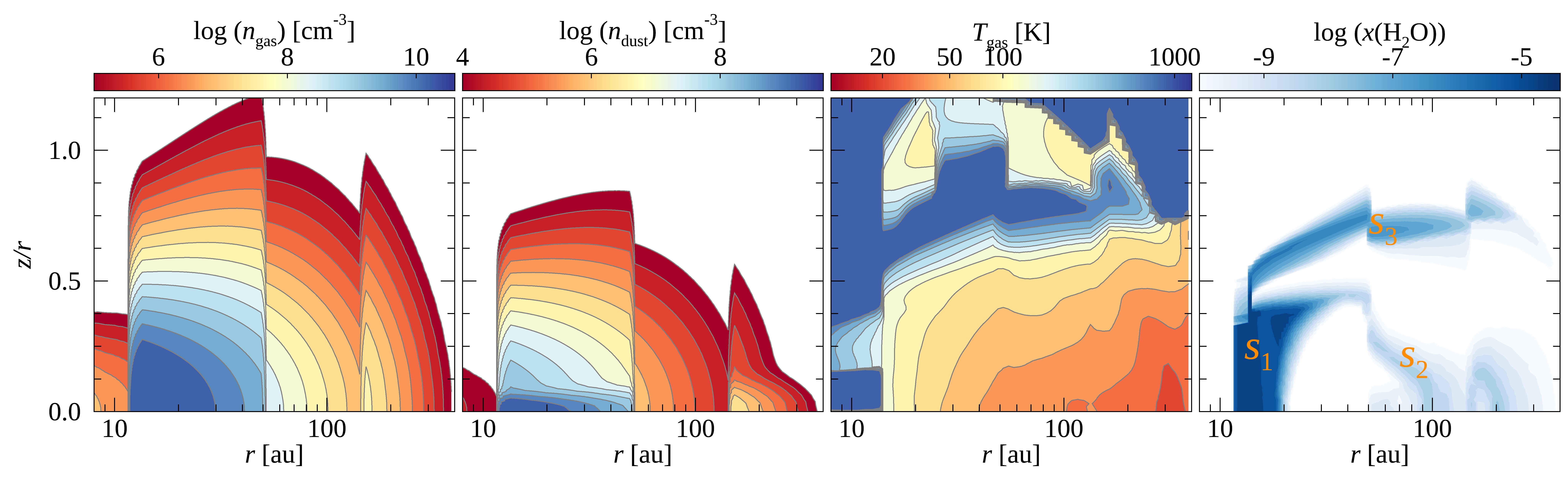}
\includegraphics[width=6cm]{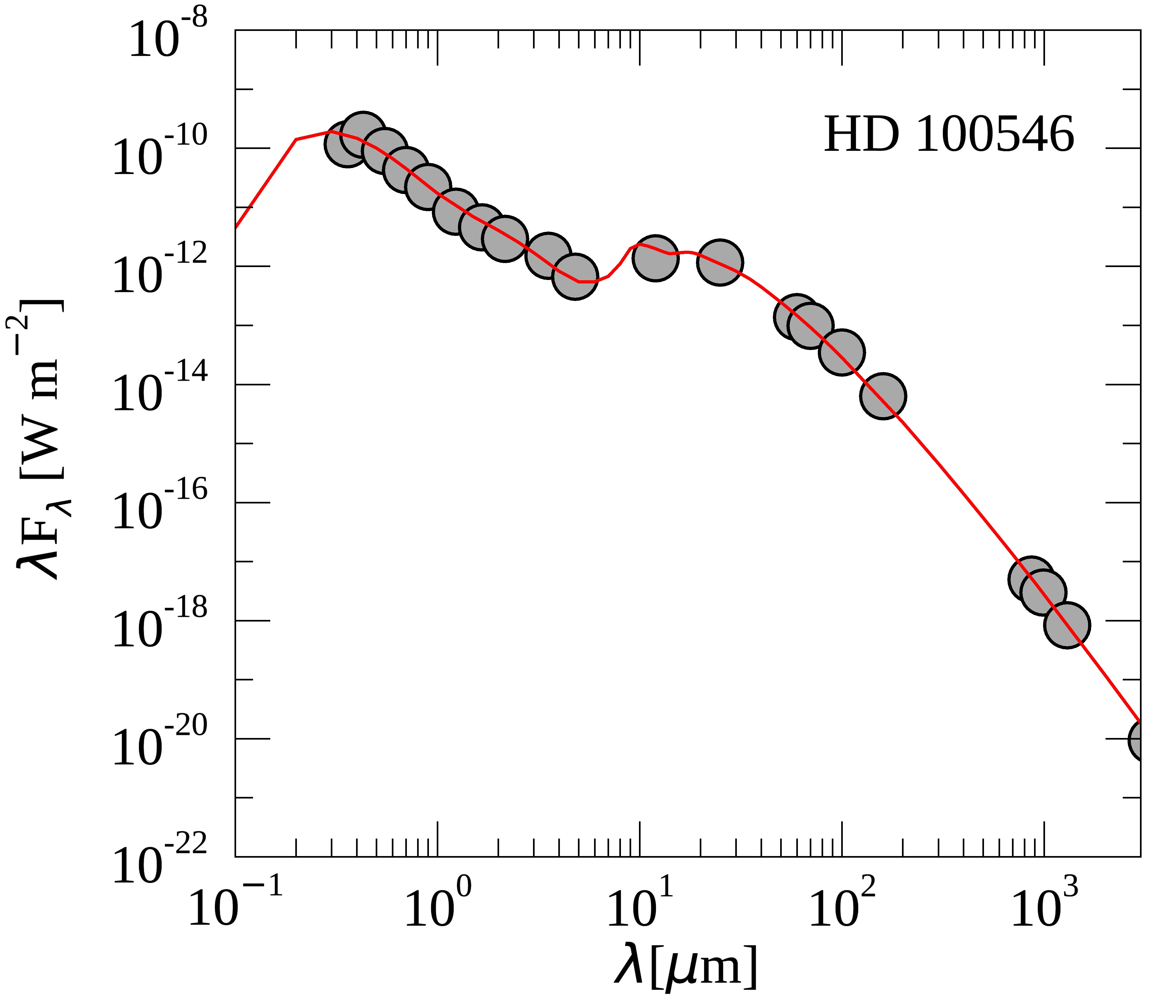}
\includegraphics[width=6cm]{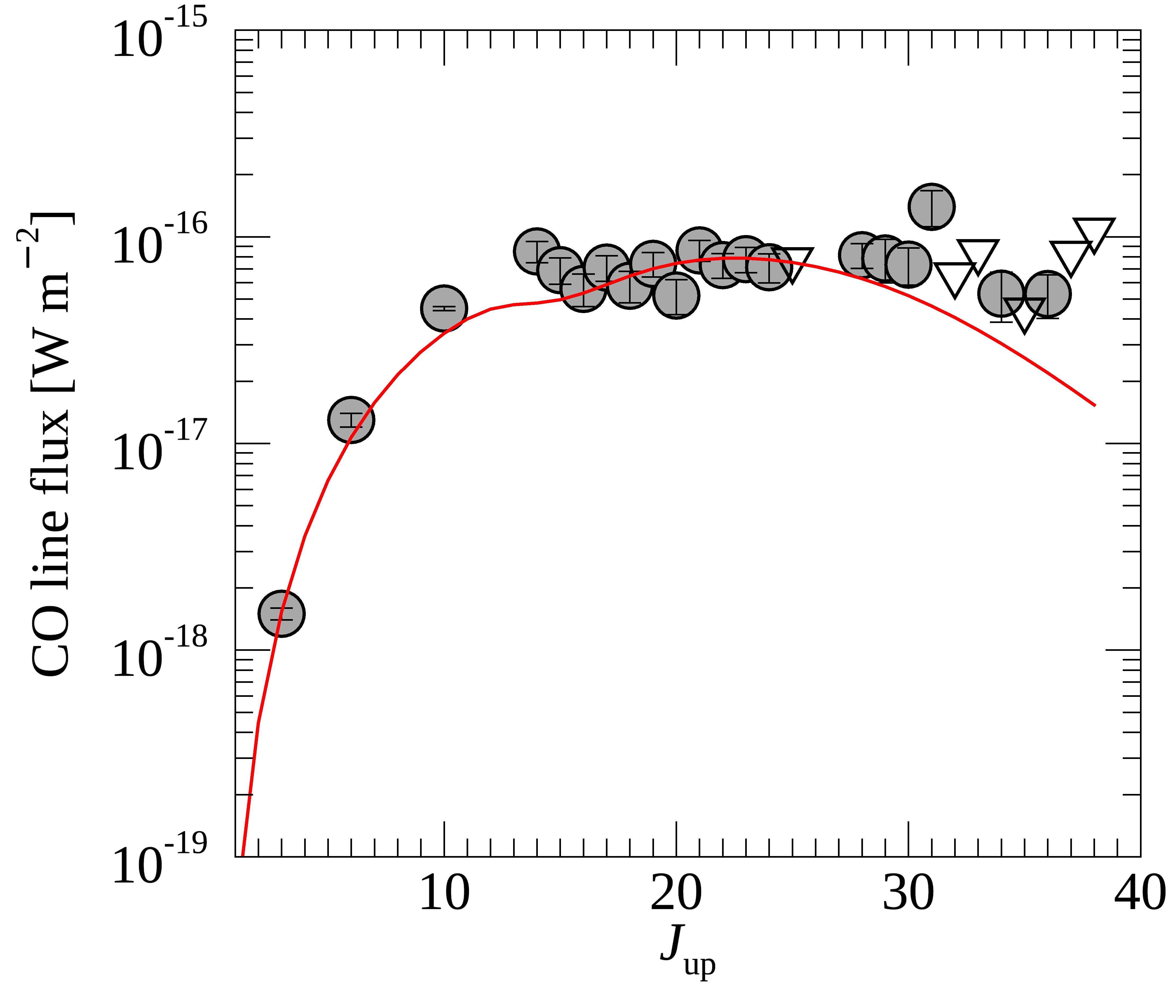}
\includegraphics[width=6cm]{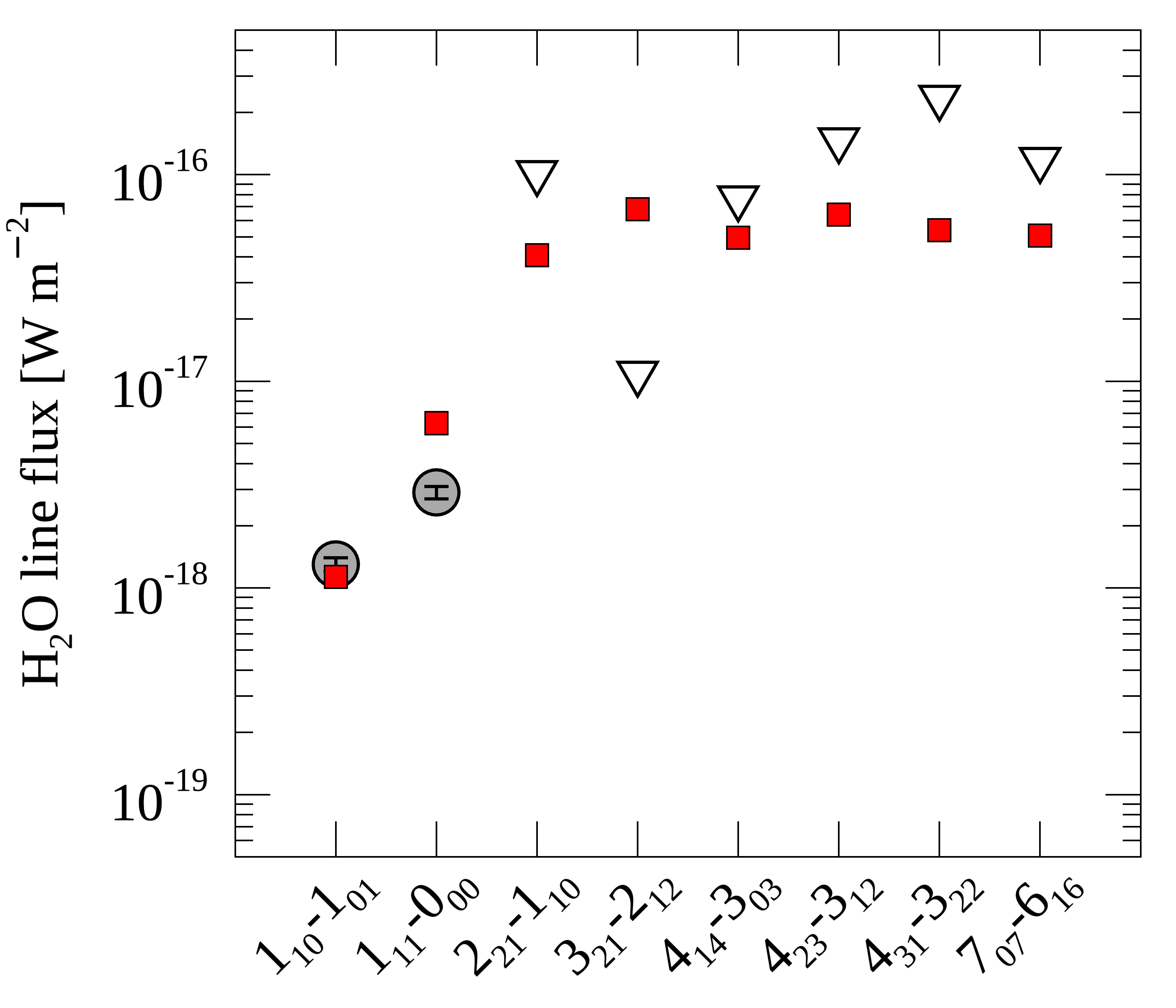}
\caption{Same as Fig.~\ref{fig:dali1} but for HD 100546.}\label{fig:dali2}
\end{figure*}

\subsubsection{The spectral energy distribution} 
The goal of this first step is to constrain the disc dust temperature and density gradients using the spectral energy distribution (SED) as an indicator. A reference physical model is obtained from the comparison of the observed SEDs with a grid of synthetic SEDs based on a grid of DALI models. 

The initial setup for the parameterisation of HD 163296 is taken from \citet{Kama20}. From the reported disc gas mass of $6.7 \times 10^{-2}\,M_{\odot}$ we derive a value of $\Sigma_\mathrm{c} = 6.8$ \gcm \ at $R_c = 125\,$ au, which was obtained by solving the mass integral. The stellar parameters have been changed to $L=17 L_{\odot}$ and $M = 2 M_{\odot}$, adapted from \cite{Vioque18}. With this setup, a first grid of simulations changing $\Sigma_{\rm c}$, $\Delta_{\rm gd}$, $h_{\rm c}$, and $\psi$ was run to determine the disc reference model. The SED is shown in Figure \ref{fig:dali1} (lower left panel).

In the case of HD 100546, the initial setup is taken from \cite{Kama16}. The initial values are $\Sigma_\mathrm{c} = 18.2$ \gcm \ at $R_c=50\,$au and $\Delta_{\mathrm{gd}} = 40$ which yield a total gas mass of $3.2 \times 10^{-2}$ $M_\odot$ and total dust mass of $8.1 \times 10^{-4}$ $M_\odot$. The \cite{Kama16} setup was derived using a distance of 97\,pc; the stellar and disc parameters were updated to account for the new distance estimate of $d = 108$ pc \citep{gaiadr3}.  
The setup includes an inner dust gap between 4\,au and 13\,au.
Recently, \citet{Fedele21} reported a new analysis of ALMA 870 $\mathrm{\mu m}$ dust continuum observations that reveals the presence of an outer gap in the disc extending from $\sim$ 40 au to $\sim$ 150\,au (hinted at by \citealt{Walsh14}) that divides the disc inner bright ring, centred at nearly 28\,au, and the disc outer faint ring, centred at $\sim$ 200 au. To reproduce the observed double-ring structure, the density structure was modified using different scaling factors ($\delta$) for the gas and the small and the large dust grains to reduce the densities in the inner dust cavity and in the outer gap: 

\begin{equation}
n_{\mathrm{i}} = 
\begin{cases}
n_{\mathrm{i}} \times \delta_{\mathrm{i,  cav}} & \text{if $R_{\rm cav, in} < R < R_{\rm cav, out}$} \\
n_{\mathrm{i}} & \text{if $R_{\rm cav, out} \leq R \leq R_{\rm gap, in}$} \\
n_{\mathrm{i}} \times \delta_{\mathrm{i,  gap}} & \text{if $R_{\rm gap, in} < R < R_{\rm gap, out}$} \\
n_{\mathrm{i}} & \text{if $r \geq R_{\rm gap, out}$} \\
\end{cases}
\label{delta}
,\end{equation}

\noindent
where $\mathrm{i} = \mathrm{gas}, \mathrm{small \ grains}, \mathrm{large \ grains}$. The values of all the 
$\delta$ factors are reported in Table \ref{tab:dali}, while the obtained gas and dust densities are shown in the top left panels of Figure \ref{fig:dali2}.
The small dust grains are assumed to be dynamically coupled to the gas, and so we impose $\delta_{\mathrm{gas}} = \delta_{\mathrm{small}}$ at every radius, while $\delta_{\mathrm{large,  cav}} = \delta_{\mathrm{large,  gap}} = 0$ to clear the inner cavity and the outer gap from the large dust grains. 
ALMA observations of CO isotopologues suggest a gas density drop of one order of magnitude in the outer gap
(Booth private communication) and so we reduced the gas density by the same amount. With this density setup, 
a grid of simulations was created varying $\Sigma_{\rm c}$, $\Delta_{\rm gd}$, $h_{\rm c}$, and $\psi$ to match the observed SED. 

\smallskip
\noindent
We note that the disc around HD 163296 is also characterised by dust gaps and rings \citep[e.g.][]{Isella16, andrews18}, but in contrast to HD 100546, the gaps are relatively narrow and their presence does not seem to affect the \water \ distribution (see discussion further below).

\subsubsection{CO rotational ladder}
In a second step, we modelled the CO rotational ladder to constrain the 2D gas temperature structure: the fluxes of the optically thick CO rotational transitions are indeed a powerful tool to trace the gas temperature throughout the radial and vertical extent of the disc \citep[][]{Fedele13b, Fedele16}. 
Observational data of the CO rotational emission in HD 163296 are taken from \cite{Fedele16}, while data of HD 100546 are taken from \cite{Meeus13, vanderwiel14, Fedele16}. The CO collisional rate coefficients are from \citet{yang10}.

To model the CO rotational ladders, a grid of DALI models was created starting from the parameters that best fit the SED and by changing parameters that mostly affect the gas structure, namely: $\Sigma_\mathrm{c}$, $\Delta_{\mathrm{gd}}$, and $\psi$. These parameters mostly change the shape of the CO rotational ladder; $\Sigma_\mathrm{c}$ and $\Delta_{\mathrm{gd}}$ are changed simultaneously to lower the total gas mass, maintaining the disc dust mass at a constant level, thus fixing the shape of the SED. 
The parameters of the reference models are given in Table~\ref{tab:dali} and the synthetic CO ladders are presented in Figures~\ref{fig:dali1} and ~\ref{fig:dali2} (middle panels) for the two discs, respectively.

\subsection{The \water \ line fluxes}
Figures \ref{fig:dali1} and \ref{fig:dali2} (lower right panel) show the flux of the \water \ rotational transitions obtained with 
the reference thermo-chemical models compared to the observational data. The DALI chemical network used here is based on the UMIST06 database \citep[][]{Woodall07} and contains ten elements (H, He, C, N, O, Mg, Si, S, Fe, and PAH), 109 species, and 1463 reactions. The chemical network is initialised with atomic abundances, with a carbon-to-oxygen abundance ratio of 0.468.

In the case of HD 163296, the reference thermo-chemical model (red squares) almost matches the low flux of the ground-state transitions but it underestimates the flux of the excited transitions by up to two orders of magnitude. On the other hand, the HD 100546 
reference model predicts a an overly strong emission in the $3_{21}-2_{12}$ transition, while it almost matches the 
flux of the ground-state transitions.

The mismatch between the observed and synthetic \water \ line fluxes cannot be due to the global physical properties, such as the density and temperature structures, as these are constrained by the SED and by the CO ladder. Similarly, the discrepancy cannot be ascribed to the global chemical composition or to the initial chemical abundances (such as the carbon/oxygen mass ratio) as in this case all lines would change accordingly. The two opposite trends of the \water \ rotational fluxes are most likely the outcome of a difference in \water \ abundance structure. To verify this hypothesis, we modelled the \water \ line fluxes adopting a parametric abundance distribution rather than a full gas-grain chemistry.

\subsubsection{Water abundance parametric distribution}\label{sec:xh2o}
Starting from the dust and gas temperatures derived above, we switched off the chemical solver and the thermal balance modules. This allows us to parametrise the \water \ abundance.
The three \water \ reservoirs can be parametrised as follows:

\begin{equation}
\mathrm {x(H_2O)} = 
\begin{cases}
x_1 & \mathrm{for} \ T_{\rm dust}  > 150\, \mathrm{K}, n_{\rm gas} > 10^9\,\mathrm{cm}^{-3} \\
x_2 & \mathrm{for} \ T_{\rm dust}  < 100\,\mathrm{K}, \mathrm{A}_{\rm V} > 3.0\,\mathrm{mag} \\
x_3 & \mathrm{for} \ T_{\rm gas} >  300\,\mathrm{K}, n_{\rm gas} > 10^6 \,\mathrm{cm}^{-3}
\end{cases}
\label{eq:xxx}
,\end{equation}

where $x_1$, $x_2$, and $x_3$ correspond to the \water \ abundance in $s_1$, $s_2$, and $s_3$, respectively.
Such a parametrisation is tuned to match the prediction of thermo-chemical models (Figures~\ref{fig:dali1} and \ref{fig:dali2}).
Once the three abundance regions had been defined, we performed a grid of DALI\ simulations varying the 
values $x_1$, $x_2$, and $x_3$. 

In parallel with the DALI  model grid, we investigated the \water \ line flux ratios by performing a set of NLTE simulations with RADEX \ \citep[][]{vanderTak07}. This is useful for checking the trend of different molecular transitions at different densities and temperatures and can be used to properly set the DALI \ model grid.

\begin{figure*}
\centering
\includegraphics[width=8cm]{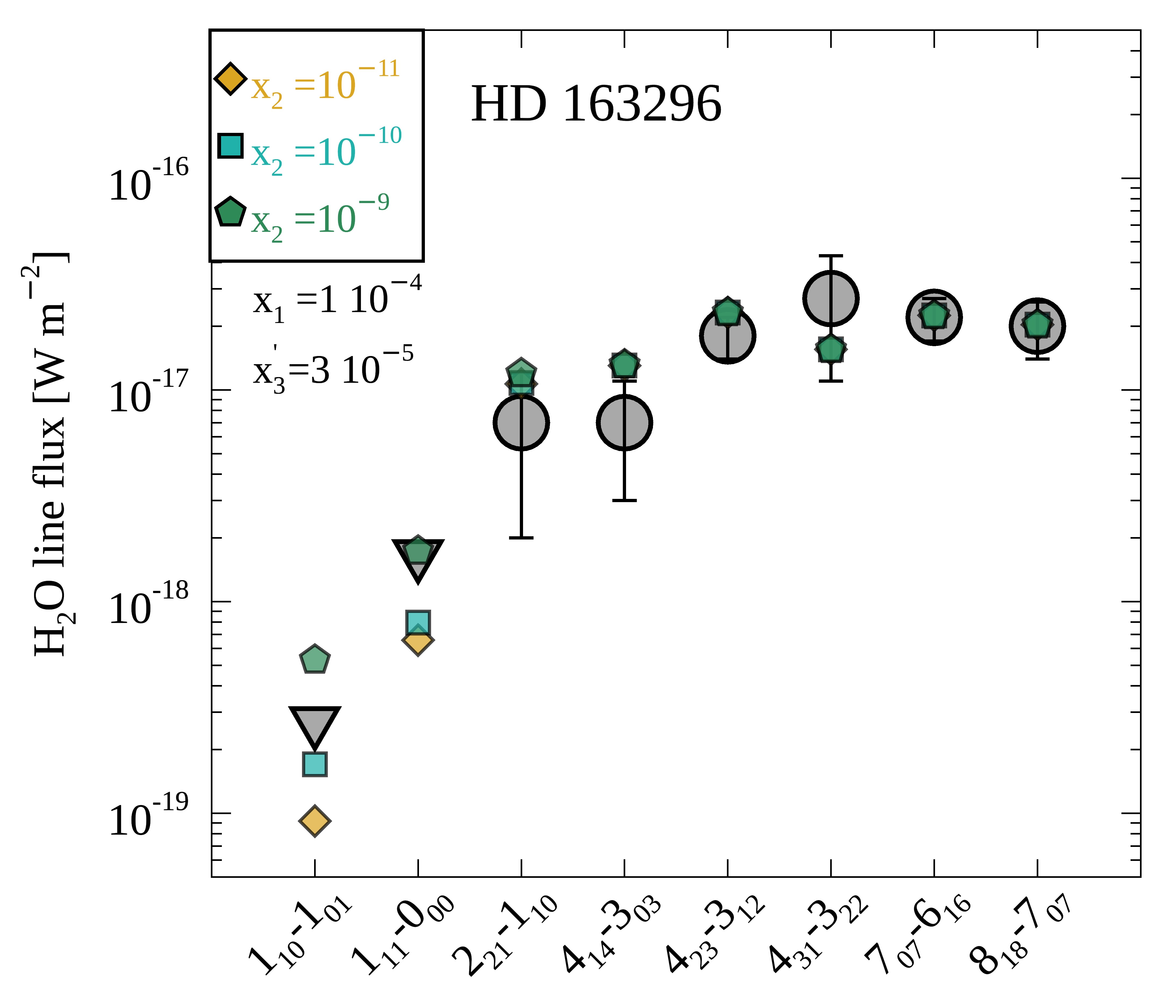}
\includegraphics[width=8cm]{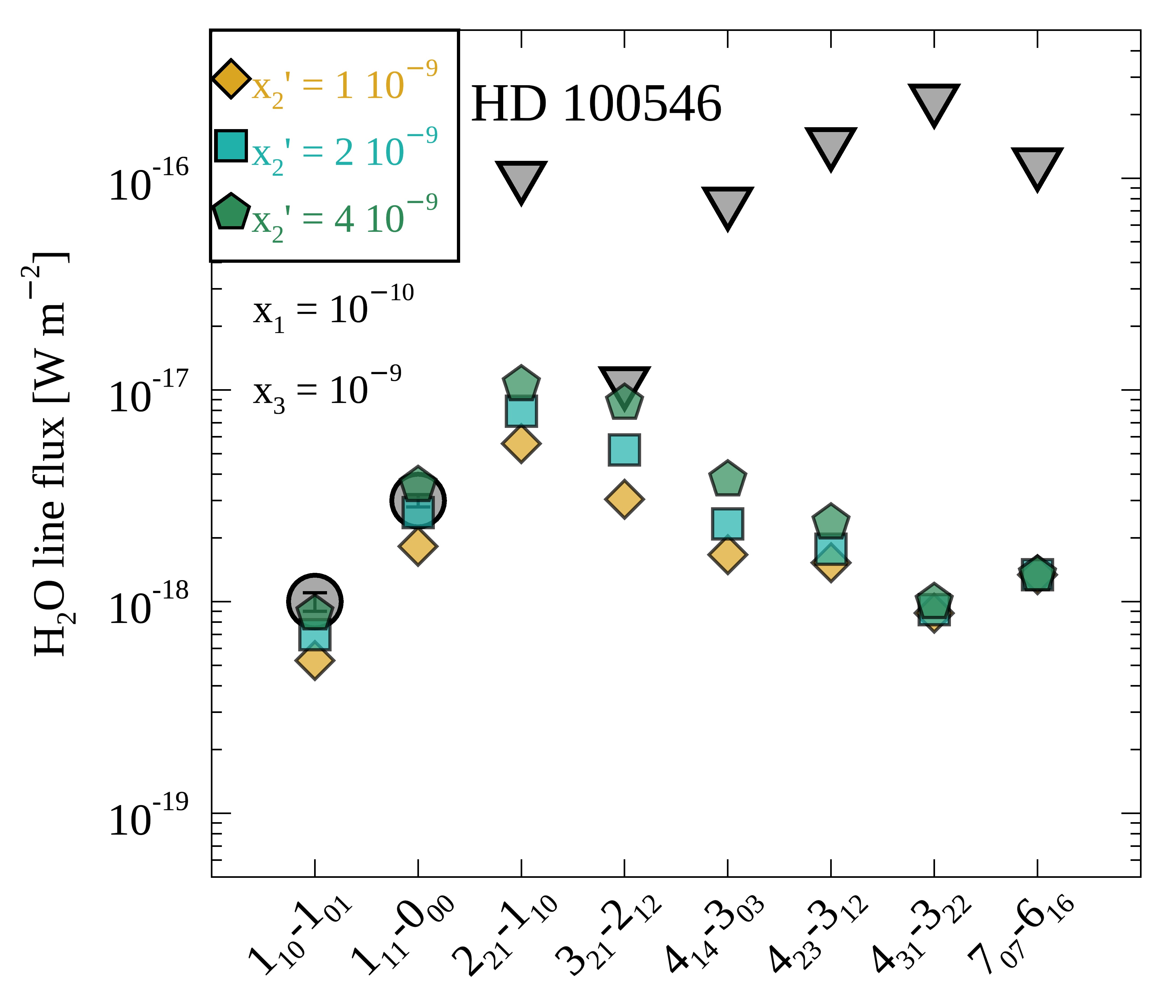}
\includegraphics[width=8cm]{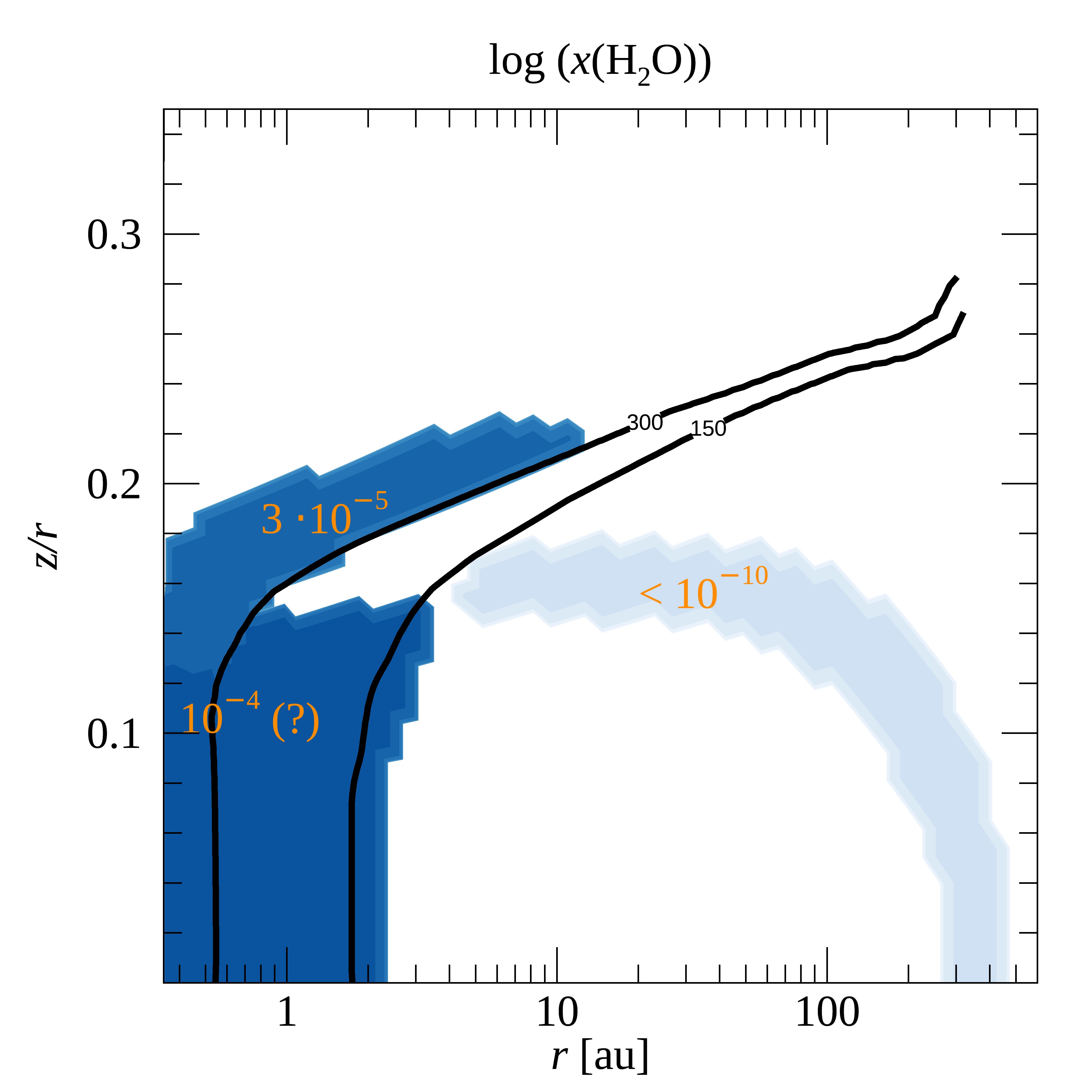}
\includegraphics[width=8cm]{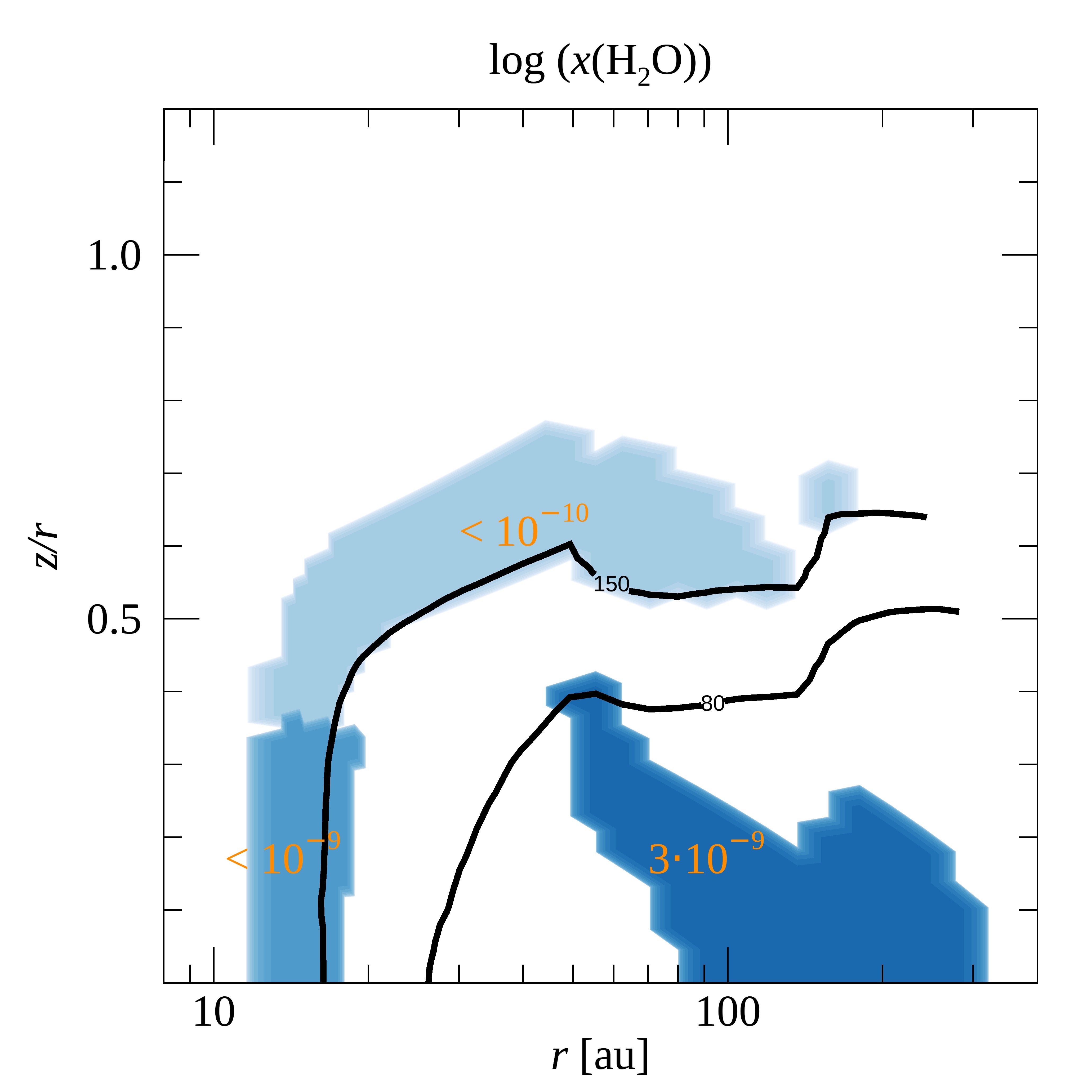}
\caption{(top) Results of the \water \ parametric models: ({\it left}) HD 163296: the models that best reproduce the observations are those with a low \water \ content in the cold reservoir ($x_3 \leq 10^{-9}$) and with no \water \ in the warm layer beyond the snow line; ({\it right}) HD 100546: models with $x_2^{\prime} \sim 2-4 \times 10^{-9}$ and low values of $x_1$ ($\leq 10^{-9}$) and $x_3$ ($\leq 10^{-10}$) match the observed flux of the ground state transitions as well as the upper limit of the high-$J$ lines. (bottom) \water \ abundance structure based on the final parametric model showing the three \water \ reservoirs.}\label{fig:final}
\end{figure*}

\subsubsection{\water \ abundance in HD 163296}
In the case of HD 163296, the non-detection of the ground-state \water \ transitions points to a low 
\water \ abundance in the cold photodesorption ($s_2$) layer. On the other hand, the 
detection of high$-J$ \water \ lines suggests a medium-to-high abundance in the warm and hot layers. 
We run an initial grid of models varying $x_1$ and $x_3$ (between $10^{-4} - 10^{-8}$) and $x_2$ ($10^{-8} - 10^{-12}$).
The results are reported in Appendix~\ref{sec:grid}.
In all cases, the models do not match the observed trend. Models with high \water \ abundance ($\gtrsim 10^{-5}$) in $s_1$ and $s_3$ almost match the flux of the most energetic transitions (though they still do not match the observations) but these models overestimate the upper limits of the low-$J$ lines independently of the value of $x_2$ (Figure ~\ref{fig:dali_grid1}). This implies that the warm layer $s_3$ contributes substantially to the ground-state transitions. Using \radex \ we investigated how the physical conditions (gas temperature and density) affect the line flux ratio between the detected mid-$J$ (e.g. $4_{14} - 3_{03}$) and undetected low-$J$ ($1_{10} - 1_{01}$) lines: the line flux ratio increases with increasing gas temperature and decreasing gas density. In particular, we find reasonable agreement with the observed flux ratio lower limit ($\gtrsim  10$) at $n_\mathrm{gas} \gtrsim 10^8$\,\density and  $T_{\rm gas} \gtrsim 250$\,K; these conditions are met only in $s_1$ and in the innermost part of the warm layer $s_3$, at $r \lesssim 10\,$au within the snow line, right above the hot inner reservoir $s_1$. 
For HD 163296, we therefore modified the parametrisation of $x_3$ as follows:


\begin{equation}
\mathrm {x(H_2O)} = 
x_3^{\prime}  \ \mathrm{for} \ T_{\rm gas} > 300\,\mathrm{K}, n_{\rm gas} > 10^8\,\mathrm{cm}^{-3}
,\end{equation}

\noindent
and we performed a new grid of models varying $x_1$, $x_2$, and $x_3^{\prime}$.  
Figure \ref{fig:final} (left panels) shows the grid results: models with $x_3^{\prime} \sim 3 \times 10^{-5}$ and $x_2 < 10^{-10}$ are in good agreement with the observed trend. We note that the line fluxes are now insensitive to the value of $x_1$, and so the actual abundance in $s_1$ is unconstrained. 

In conclusion, we find that HD 163296 hosts a H2O-rich inner disc inside the \water \ snow line ($r \sim 5$ AU), while the disc outer region is \water-poor. The high-$J$ \water \ lines detected by PACS require a hot inner reservoir that includes a density of  higher than $10^8$ \gasdensity with a \water\  abundance of $x_1 = 10^{-4}$. On the contrary, from the non-detection of the low-$J$ \water \ lines, we constrain the \water \ abundance in the photodesorption layer ---defined for 3 $< A_{\mathrm{V}} <$ 5 outside the disc snow line--- to be $x_2 < 10^{-9}$ and the \water \ abundance in the disc warm layer to be negligible. Based on the revised parametrisation, we computed the synthetic spectrum of \water \ with DALI. The results are reported in Appendix~\ref{sec:synthetic}.

\subsubsection{\water \ abundance in HD 100546}
In the case of HD 100546, the detection of the ground-state \water \ lines implies a contribution from the
cold photodesoption layer ($s_2$). Contrary to HD 163296, the high-$J$
transitions in the PACS range remain undetected with flux upper limits of the order of a few 
$10^{-17}$\,W\,m$^{-2}$ \citep{Meeus12, Fedele13a}. The deep PACS observation of the $3_{21}-2_{12}$
transition at 75.38\,\micron \ is an order of magnitude more sensitive than the previous observations and yet
the line is not detected. These non-detections suggest a low \water \ abundance in the warm layer of HD 
100546. 

\noindent
We ran an initial grid of parametric models, varying the \water \ abundance in the three reservoirs in the ranges $10^{-8} - 10^{-12}$ (in $s_1$ and $s_3$) and $10^{-8}-10^{-10}$ (in $s_2$). The results of the model grid are shown in Figure~\ref{fig:dali_grid2}. As expected, the fluxes of the lowest transitions (up to the $3_{21}-2_{12}$) are mostly dominated by the abundance in the cold reservoir, while both $s_1$ and $s_3$ contribute to the flux of the the high-$J$ lines.  

Of key importance is the upper limit of the $3_{21}-2_{12}$ transition as all the models in the initial grid overestimate the upper limit (Figure~\ref{fig:dali_grid2}). Using \radex, \ we inspected how the line flux ratios (e.g. $3_{21} - 2_{12} / 1_{10} - 1_{01} \lesssim 8$) vary with gas temperature and density, and \water \ column density. The only way to match the observed flux ratio is to lower the gas temperature ($\lesssim 80\,$K) in the line-emitting region. This imposes a very low abundance in $s_3$ (otherwise the upper level is easily populated) as its upper layers are known to be warm from the analysis of the CO ladder. Interestingly, the independent analysis of the line velocity profile of the HIFI \water \ spectra allows us to pose a robust constraint on the line-emitting region in the cold reservoir: as reported by \citet{vanDishoeck21}, the ground-state \water \ emission comes from $\sim 40$\,au outward. Based on this observational finding, we slightly modified the parametrisation of $s_2$ to be limited beyond 40 au (equivalent to $T_{\rm dust} < 80\,$K):

\begin{equation}
\mathrm {x(H_2O)} = 
x_2^{\prime}  \ \mathrm{for} \ T_{\rm dust} < 80\,\mathrm{K}, \mathrm{A_{\rm V}} > 3.0\,\mathrm{mag}
.\end{equation}

\begin{figure}
    \centering
    \includegraphics[width=8cm]{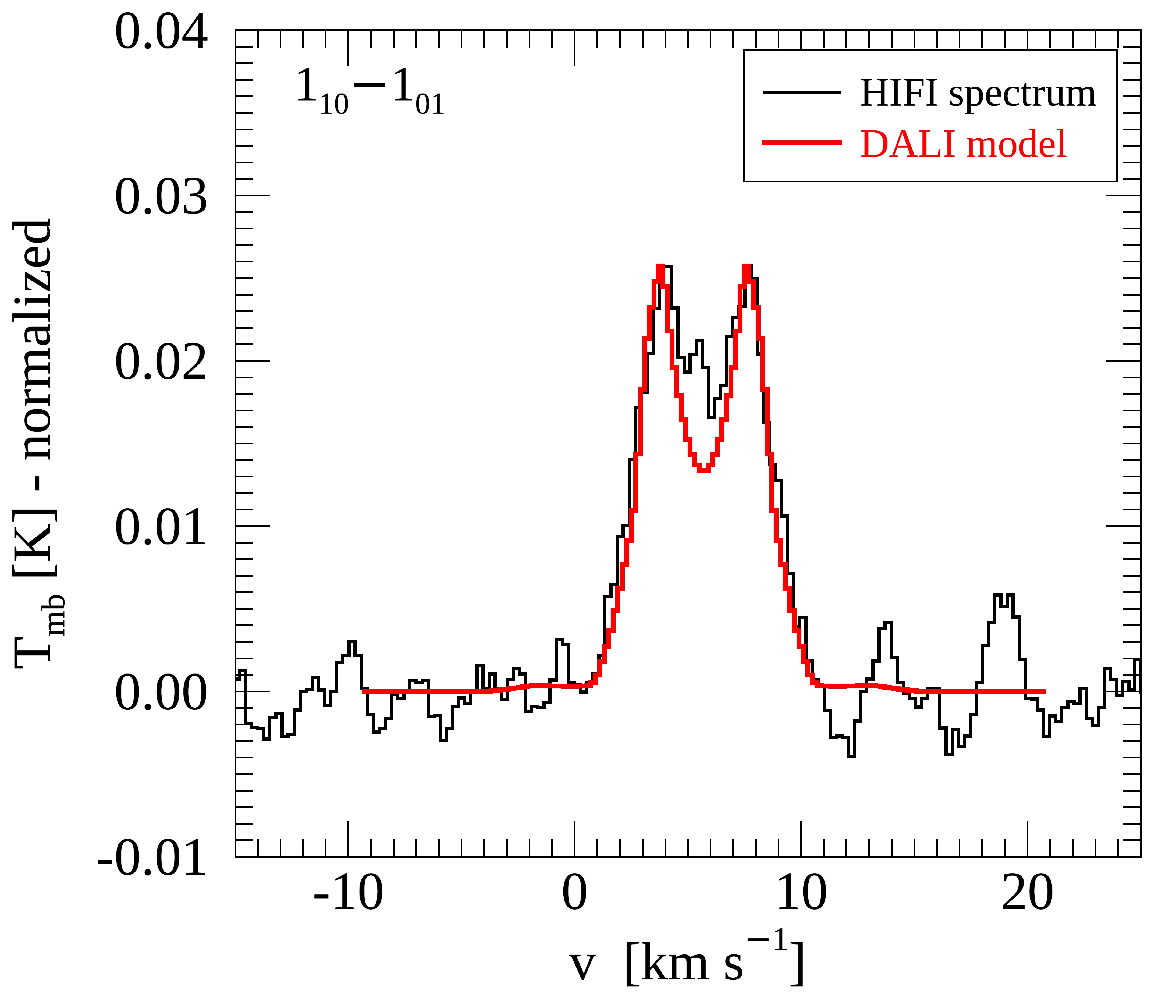}
    \includegraphics[width=8cm]{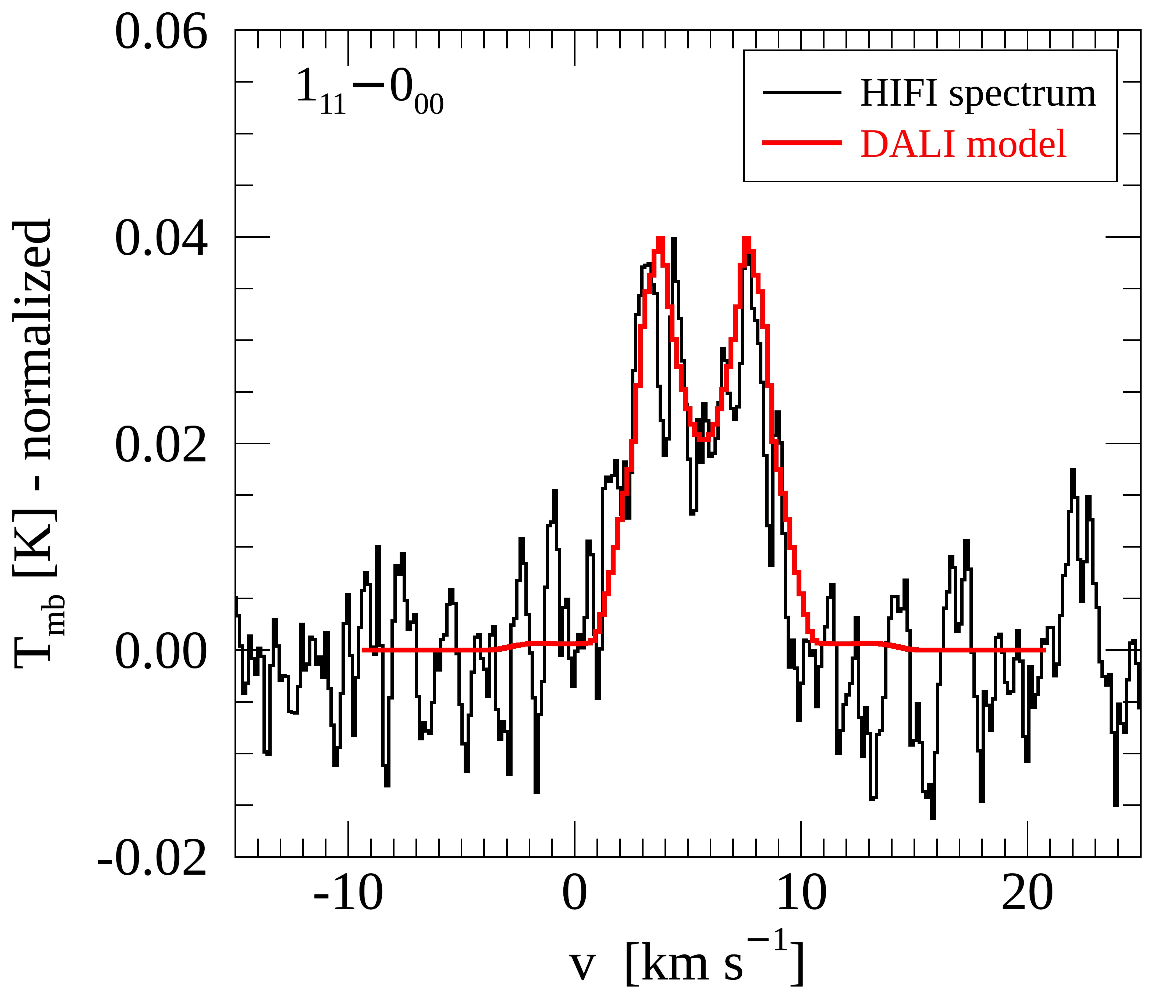}
    \caption{Velocity profile of the ground-state \water \ lines in HD 100546 and comparison with the DALI model spectrum. The spectra are continuum-subtracted and normalised. The DALI model nicely reproduces the observed profile of both transitions.}
    \label{fig:profile}
\end{figure}

\noindent
With this assumption, the models with $x_2^{\prime} \sim 3 \times 10^{-9}$ and low values of $x_1$ ($\leq 10^{-9}$) and $x_3$ ($\leq 10^{-10}$) are in good agreement with the observational trend (Figure~\ref{fig:final}, right). 
Figure~\ref{fig:profile} shows the observed and synthetic line velocity profile of the ground-state transitions. The DALI synthetic spectra nicely reproduce the observed line profile.
Also, in this case we computed the \water \ spectrum based on the parametrised distribution and the result is shown in Appendix.~\ref{sec:synthetic}.

\begin{figure*}
\centering
\includegraphics[width=9cm]{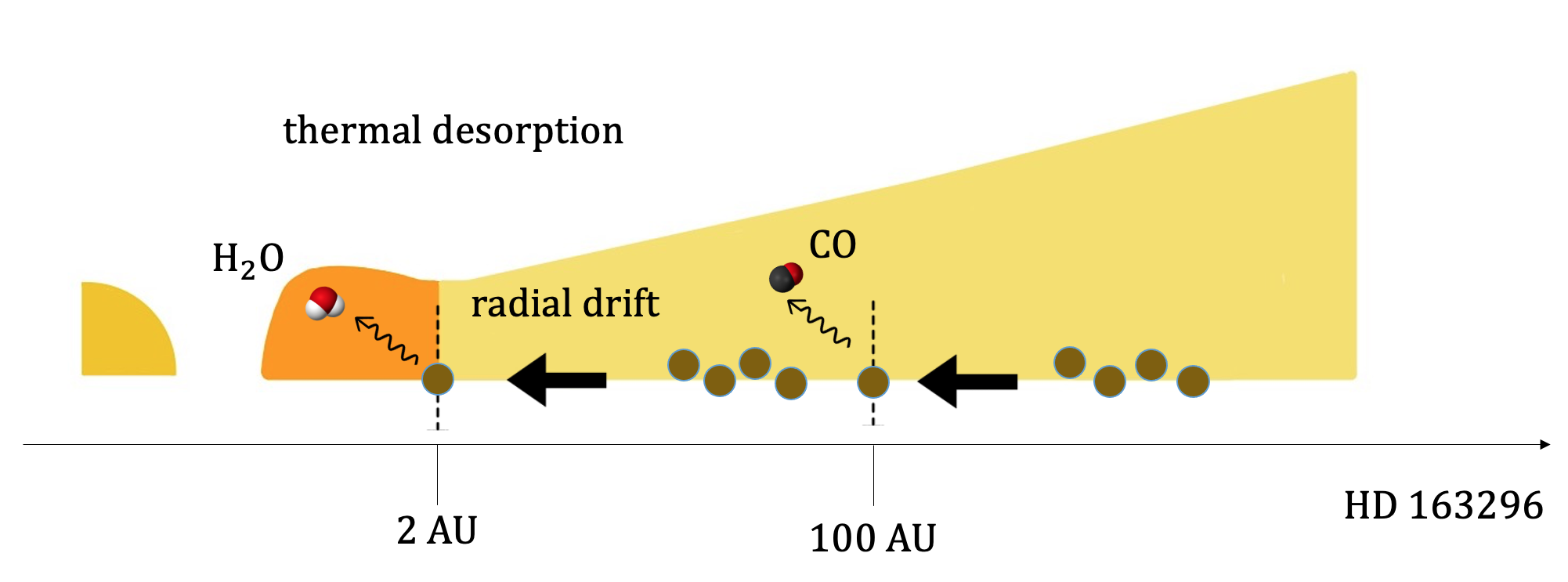}
\includegraphics[width=8cm]{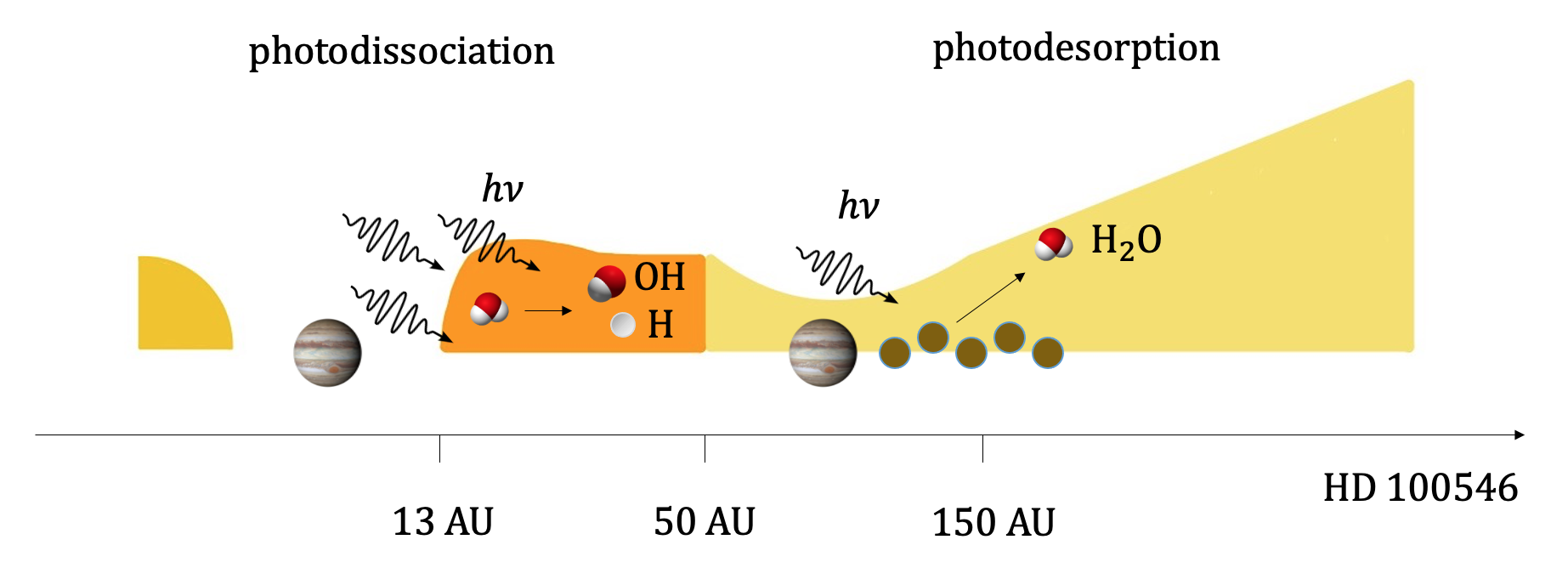}
\includegraphics[width=8cm]{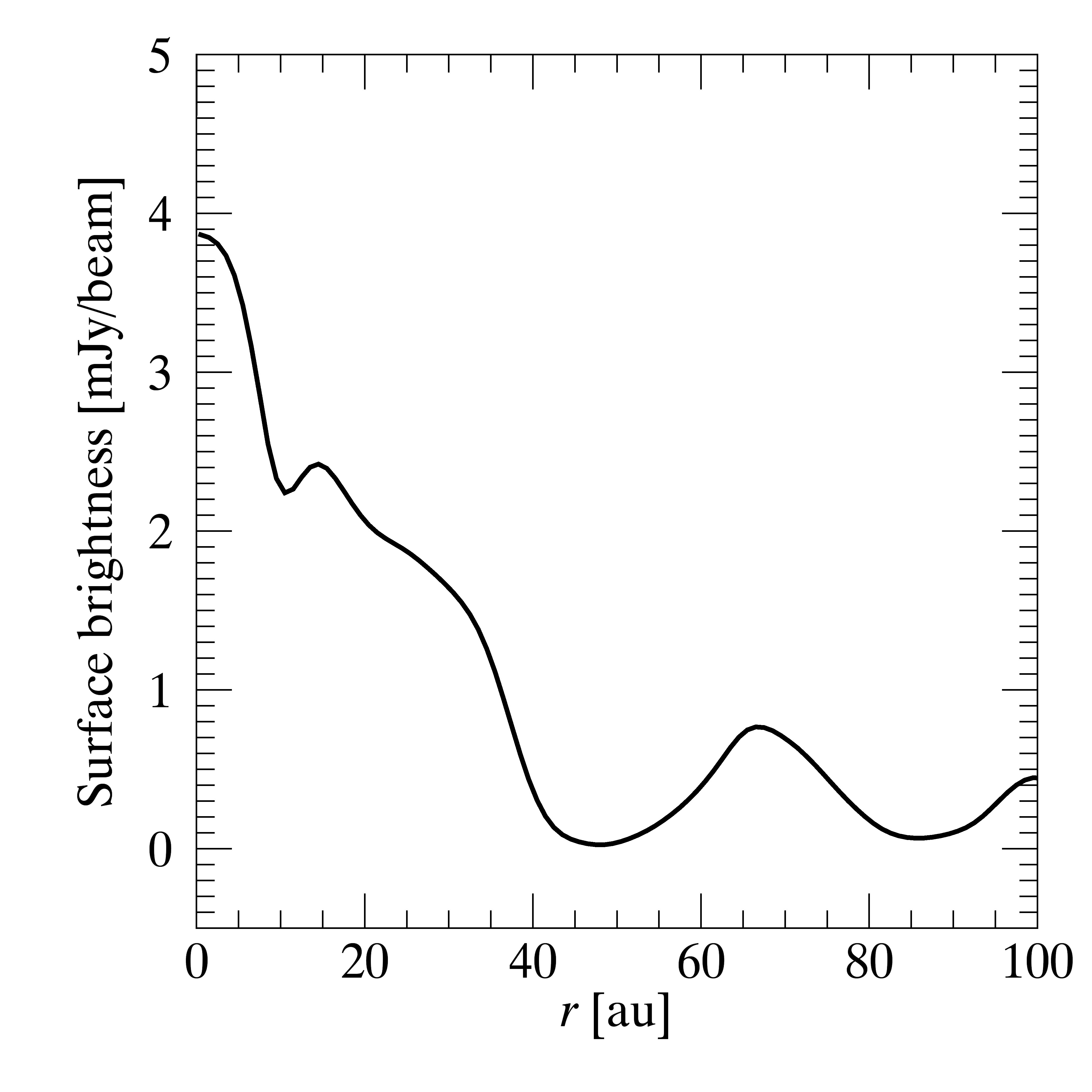}
\includegraphics[width=8cm]{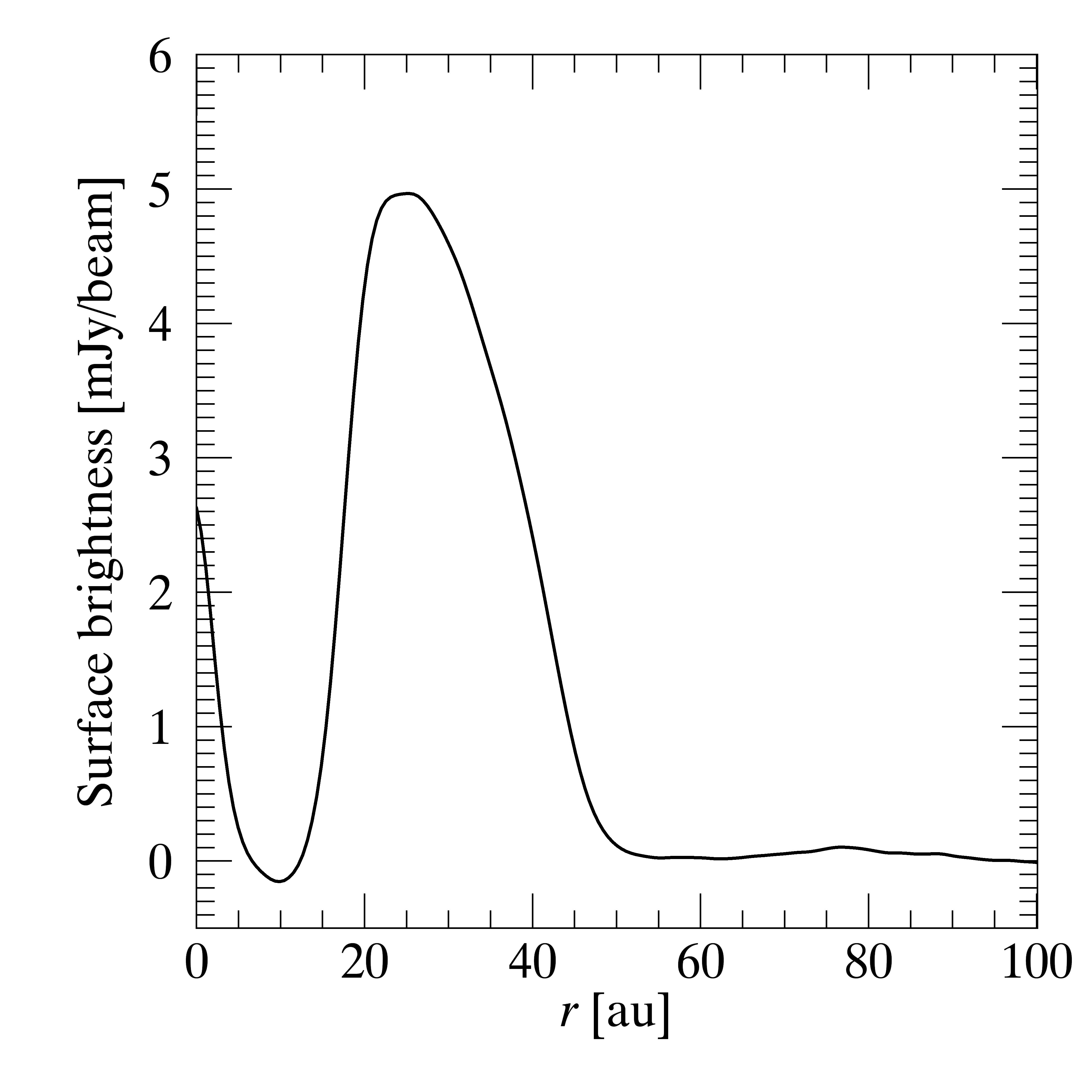}
\includegraphics[width=8cm]{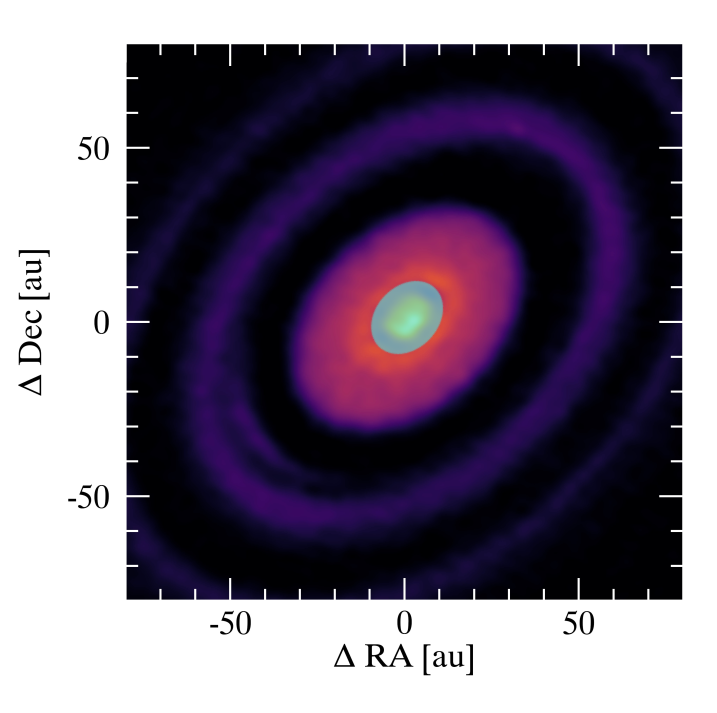}
\includegraphics[width=8cm]{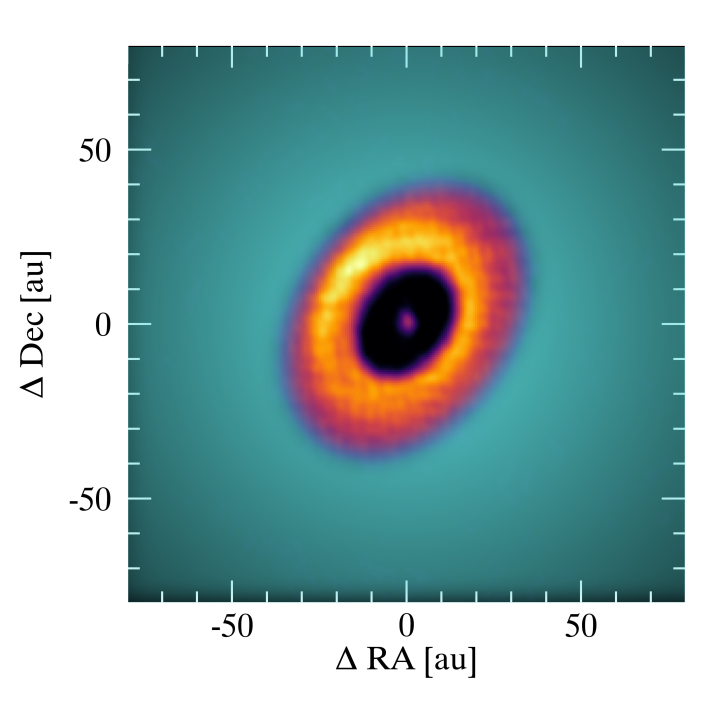}
\caption{(top) Schematic of the \water \ distribution in the two discs.  (middle) Surface brightness profile (azimuthally averaged) of the millimetre dust continuum at 1.2\,mm for HD 163296 \citep[from DSHARP survey,][]{andrews18, huang18a} and at 0.87\,mm for HD 100546 \citep{Pineda19, Fedele21} in the inner 100\,au. (bottom) Water-emitting region (cyan) overlaid on the ALMA dust continuum image. Notably in the case of HD 163296, the size of the \water -emitting region corresponds to the narrow dust gap at 10\,au. This may be due to dust growth at the border of the snow line.}\label{fig:schematic}
\end{figure*}

\section{Discussion and Conclusion}
The dichotomy of the gas-phase \water \ abundance structure in HD 163296 and HD 100546 can be explained through the dynamical history of the two discs, which influences their chemistry.

\citet{Banzatti20} proposed a link between the inner disc chemistry and the migration of solid pebbles for \water \ while \cite{Booth19} and \cite{Zhang21} speculate that the enhanced CO column density in the disc inner 100 au of HD 163296 is due to the inward drift of icy dust particles that release the CO to the gas phase when they cross the CO frost line at $r \approx$ 100\,au.  A similar mechanism might explain the high \water \ abundance within the disc snow line. Notably, a narrow dust gap is visible in the ALMA dust continuum at $r=10\,au$ (Figure~\ref{fig:schematic}). This spatial coincidence might be due to dust growth at the border of the snow line as suggested by \citet{Zhang15} in the case of HL Tau. Further analysis is needed to verify such a hypothesis. In this regard, \citet{notsu19} find that the we should observe a change in the millimetre dust opacity at the snow line position. 

In HD 100546,  the \water \ emission region overlaps with a disc outer gap between $r \sim 40$ au and $r \sim 150$ au detected with ALMA at 870~\micron ~\citep{Fedele21}. Notably, this is spatially coincident with the 3~\micron ~\water ~ice absorption band detected by \citet{Honda16} that extends from $r \sim 40$ au to $r \sim 120$ au. The high \water \ abundance in the photodesorption layer can be explained through the spatial coincidence of the dust gap and the \water -emitting region: the presence of the outer disc gap lets the UV stellar photons penetrate to the colder layers to make \water \ ice sublimate through photodesorption. This is shown in the schematics presented in Figure~\ref{fig:schematic}. This scenario can also explain the origin of the crystalline forsterite emission at 69$\mu$m: the blackbody fitting suggests that the emission arise from dust grains at $\sim$ 70\,K at radial distances $r > $ 50\,au \citep{Sturm10}. As for the \water \ ice, the crystalline forsterite is directly illuminated by the stellar UV photons that enter in the dust gap. 

Interestingly, \cite{Booth21} report the detection of gas-phase \methanol \ towards HD 100546: contrary to \water , the \methanol \ emission peaks in the inner 60\,au of the disc and extends out to nearly 300\,au. According to \citet{Booth21}, the \methanol \ emission is due to desorption of the  primordial ice inherited from the cold dark cloud: in particular, the bulk of the emission from the inner disc is due to thermal desorption mostly from the inner cavity at 13\,au, while the spatially extended emission is due to photodesorption, similarly to the scenario proposed here for \water .

It is worth noting that the only other disc with a detection of cold gas-phase \water, TW Hya \citep{Hogerheijde11}, also shows \methanol \ emission \citep{Walsh16}. 

The low \water \ abundance in the outer warm molecular layer in both HD 163296 and HD 100546 implies an O-poor molecular layer compared to the standard thermochemical predictions. This is in agreement with the findings of low gas-phase CO abundances and large carbon/oxygen mass ratios observed in several discs  \citep[e.g.][]{Bergin10, Favre13, McClure16, Kama16, Miotello17, Semenov18}.  

In both cases, oxygen in the outer disc is incorporated into dust particles settled to the midplane \citep[e.g.][]{Krijt20}. In the case of HD 163296, the radial migration of icy pebbles to the inner disc decreases the \water ~content in the outer disc and enhances the gas-phase reservoir inside the \water ~snow line. On the other hand,  in HD 100546,  the presence of the outer wide gap (see Figure~\ref{fig:schematic} bottom panels) prevents or slows down the radial drift of the ice dust grains towards the high-temperature region, resulting in a (relatively) high gas-phase \water ~abundance in the outer disc. Compared to the wide and deep gap of HD 100546, the three outer gaps observed in HD 163296 (Figure~\ref{fig:schematic}) are not wide and deep enough to stop the inward drift of icy grains. Our results show how the chemistry of discs is strictly connected to the dynamical history of the system and to the presence of giant protoplanets.

\begin{acknowledgements}
DF acknowledges the support of the Italian National Institute of Astrophysics (INAF) through the INAF Mainstream projects ARIEL and the ``Astrochemical Link between Circumstellar Disks and Planets", ``Protoplanetary Disks Seen through the Eyes of New- generation Instruments" and by the PRIN-INAF 2019 Planetary Systems At Early Ages (PLATEA). This project has received funding from the European Union’s Horizon 2020 research and innovation programme under the Marie Skłodowska-Curie grant agreement No 823823 (DUSTBUSTERS).
This work has made use of data from the European Space Agency (ESA) mission
{\it Gaia} (\url{https://www.cosmos.esa.int/gaia}), processed by the {\it Gaia}
Data Processing and Analysis Consortium (DPAC,
\url{https://www.cosmos.esa.int/web/gaia/dpac/consortium}). Funding for the DPAC
has been provided by national institutions, in particular the institutions
participating in the {\it Gaia} Multilateral Agreement.
HIFI was designed and built by a consortium of institutes and university departments
from across Europe, Canada and the US under the leadership of SRON Netherlands Institute
for Space Research, Groningen, The Netherlands with major contributions from Germany, 
France and the US. Consortium members are: Canada: CSA, U. Waterloo; 
France: CESR, LAB, LERMA,IRAM; 
Germany: KOSMA, MPIfR, MPS; Ireland, NUI Maynooth; 
Italy: ASI,IFSI-INAF,  Arcetri-INAF;  
Netherlands:  SRON,  TUD;  
Poland:  CAMK,  CBK; 
Spain: Observatorio  Astronómico  Nacional  (IGN),  Centro  de  Astrobiologıa(CSIC-INTA);  
Sweden:  Chalmers  University  of  Technology  -  MC2,  RSS  \& GARD,  Onsala  Space  Observatory,  Swedish  National  Space  Board,  Stock-holm University - Stockholm Observatory; 
Switzerland: ETH Zürich, FHNW; 
USA:  Caltech,  JPL,  NHSC.  
PACS has been developed by a consortium of institutes led by MPE (Germany) and including UVIE (Austria);  
KUL, CSL,IMEC (Belgium); CEA, OAMP (France); MPIA (Germany); IFSI, OAP/OAT, OAA/CAISMI, LENS, SISSA (Italy); IAC (Spain). This development has been supported  by  the  funding  agencies  BMVIT  (Austria),  ESA-PRODEX  (Bel-gium), CEA/CNES (France), DLR (Germany), ASI (Italy), and CICYT/MCYT(Spain). 

\end{acknowledgements}


\newpage 
\begin{appendix}

\section{\water \ model grid}\label{sec:grid}
\begin{figure}[!ht]
\centering
\includegraphics[width=8cm]{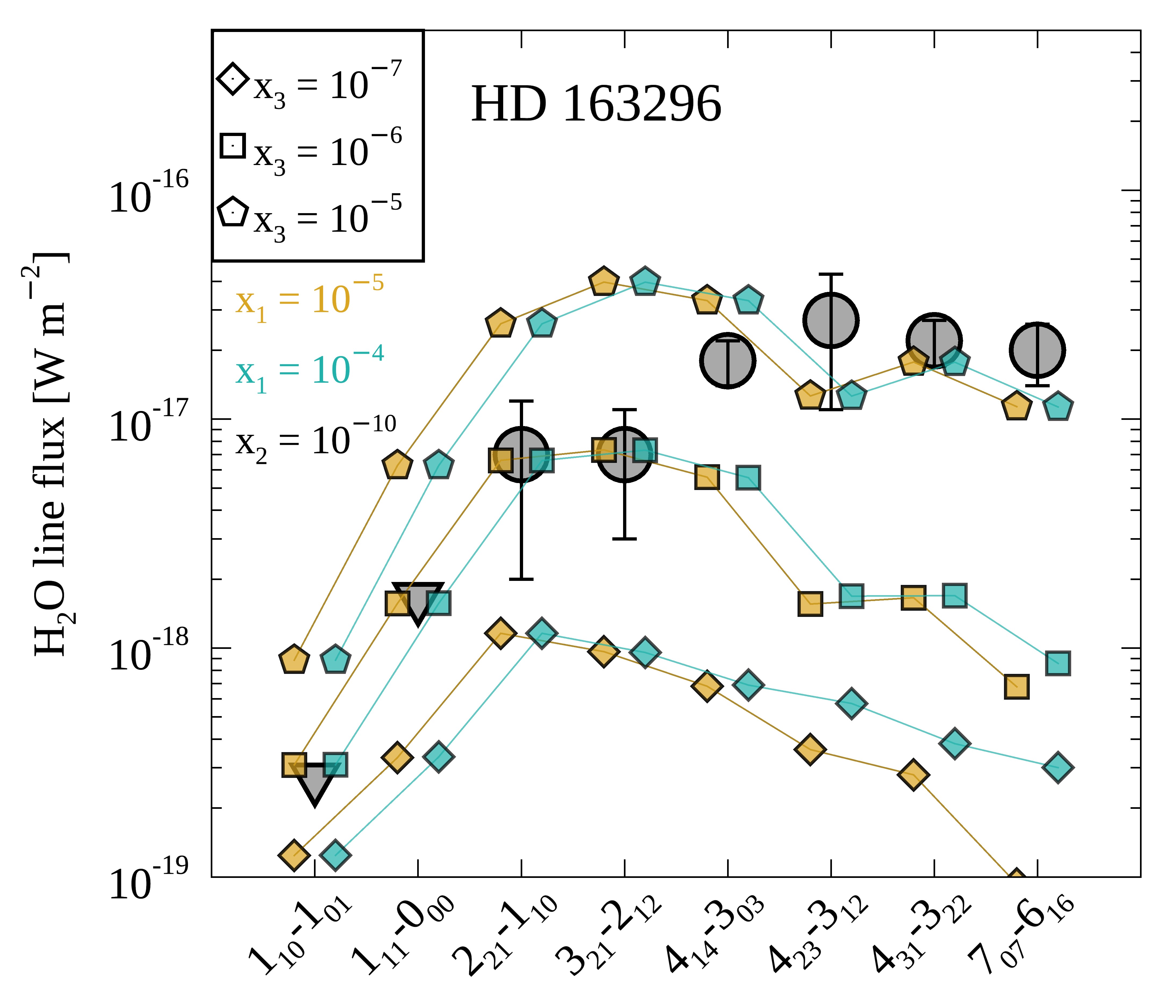}
\includegraphics[width=8cm]{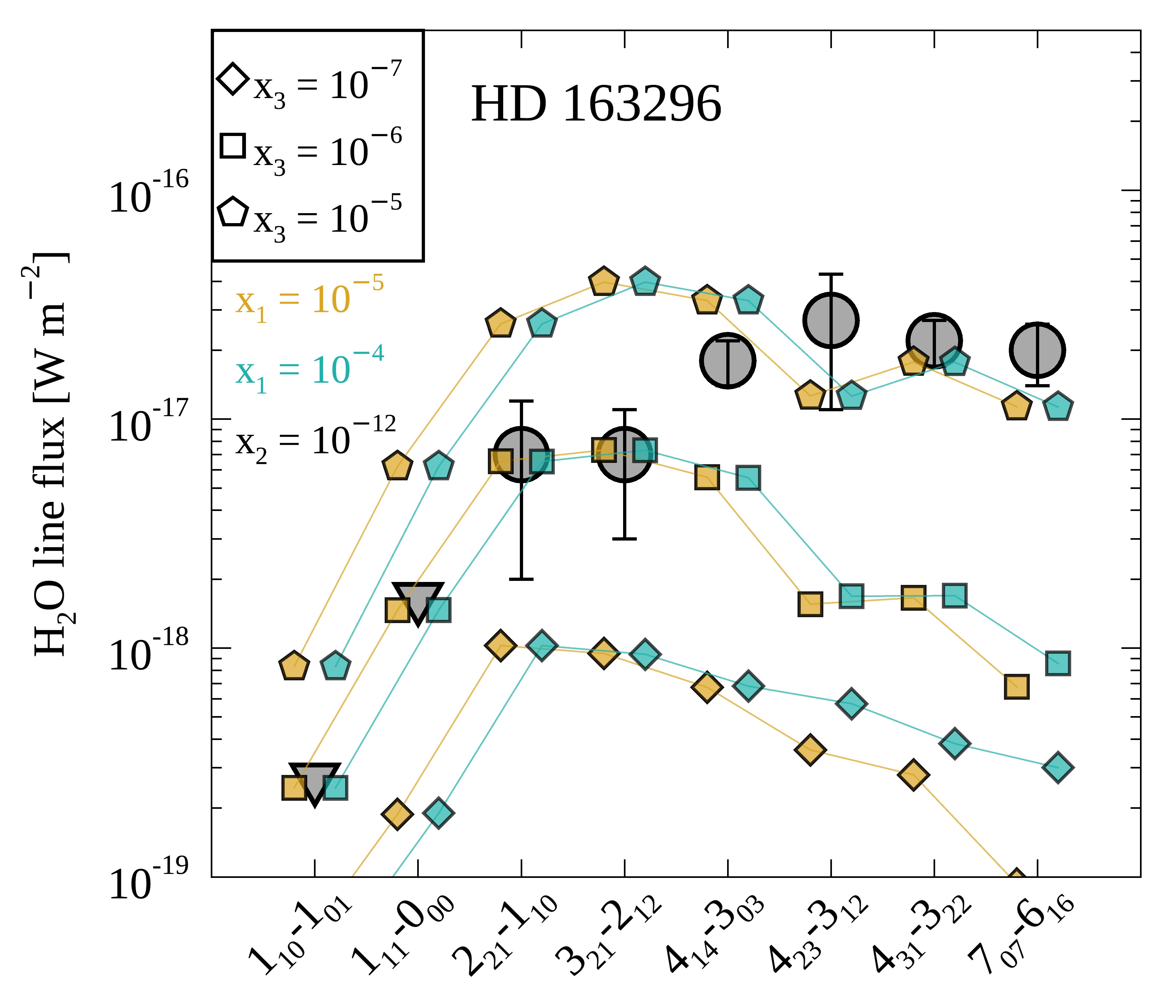}
\caption{Results of the parametric model grid for HD 163296.}\label{fig:dali_grid1}
\end{figure}

\begin{figure}[!ht]
\centering
\includegraphics[width=8cm]{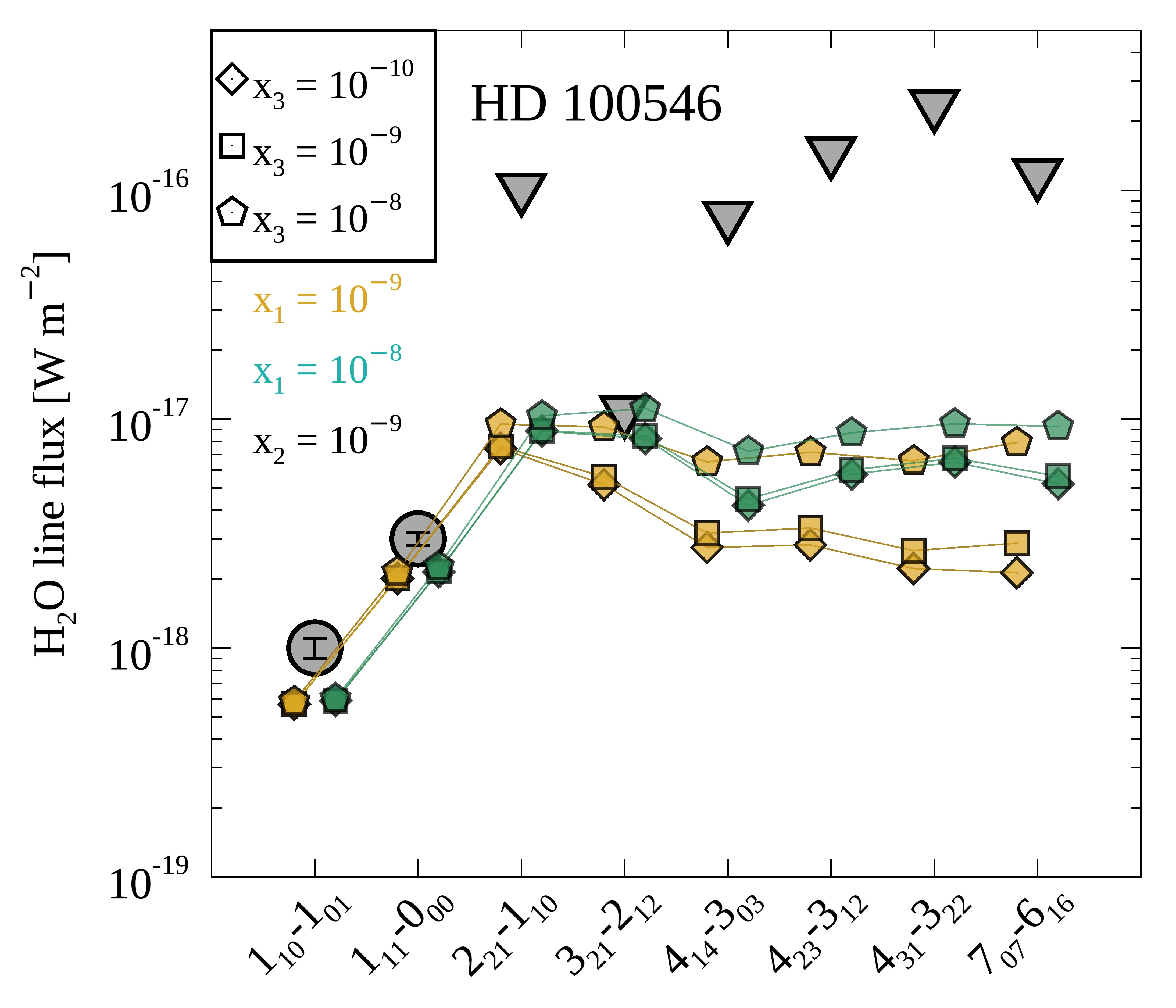}
\includegraphics[width=8cm]{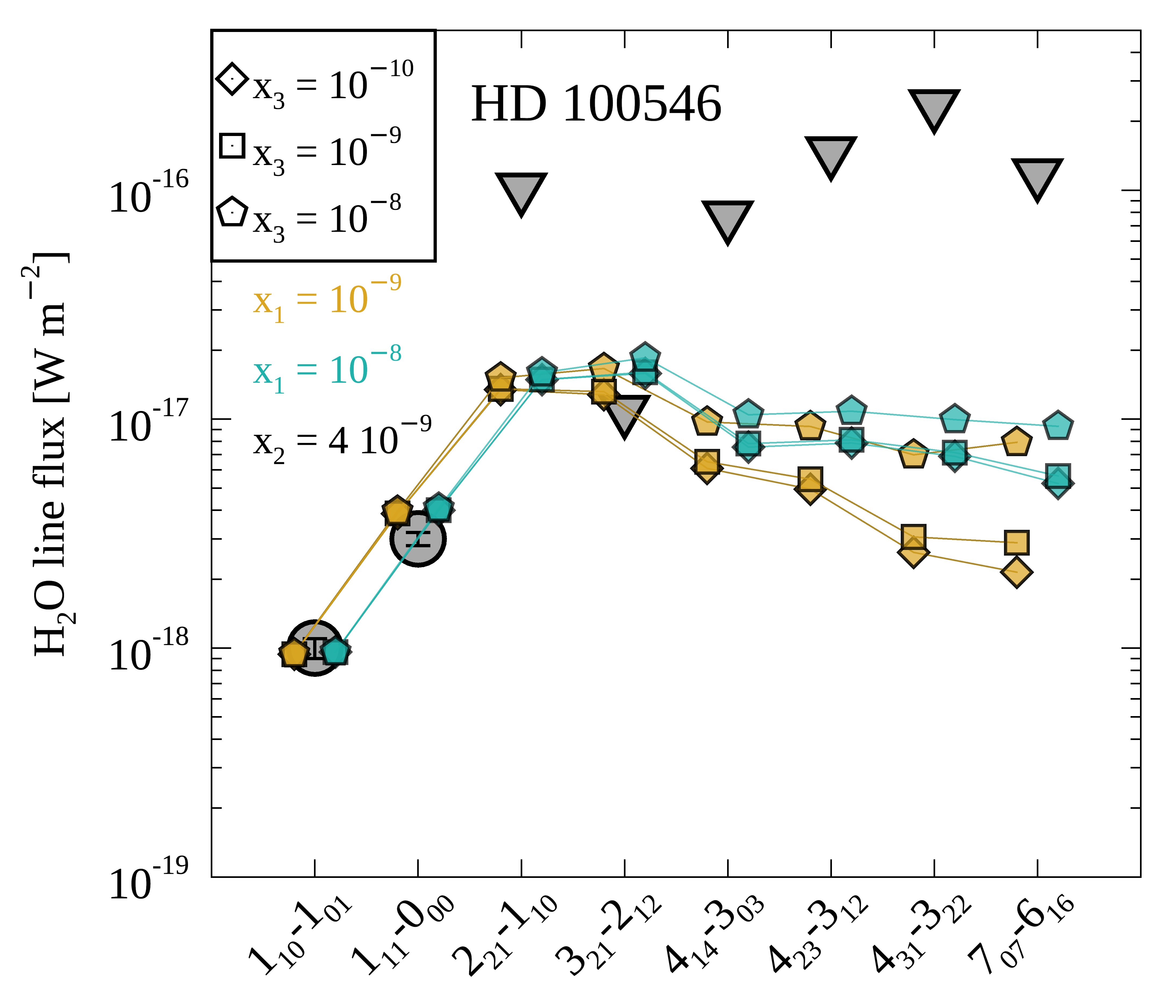}
\caption{Same as Figure~\ref{fig:dali_grid1} but for HD 100546.}\label{fig:dali_grid2}
\end{figure}

The results of the parametric model grid for HD 163296 and HD 100546 are shown in Figures~\ref{fig:dali_grid1} and \ref{fig:dali_grid2}. The models shown here refer to the original definition of the \water \ reservoir (Equation~\ref{eq:xxx}). Only a portion of the grid is shown here.

In the case of HD 163296, none of the models are able to reproduce the observed trend: on the one hand, the high-$J$ line requires an high abundance ($\gtrsim 10^{-5}$, Figure~\ref{fig:dali_grid1}) in the warm layer ($s_3$) but these models overestimate the upper limit of the two ground-state transitions. Lowering the value of $x_2$ does not help.  

Figure~\ref{fig:dali_grid2} shows a subset of the model grid for two different values of $x_2$, 10$^{-9}$ and $4 \times 10^{-9}$. In the first case, the models underestimate the flux of the ground-state transitions. Increasing the \water \ abundance in $s_2$ is better suited to fit the flux of the cold transitions but note that these models fail to reproduce the upper limit to the $3_{21} - 2_{12}$ transition.

\section{Synthetic \water \ spectrum}\label{sec:synthetic}
\begin{figure}
    \centering
    \includegraphics[width=8cm]{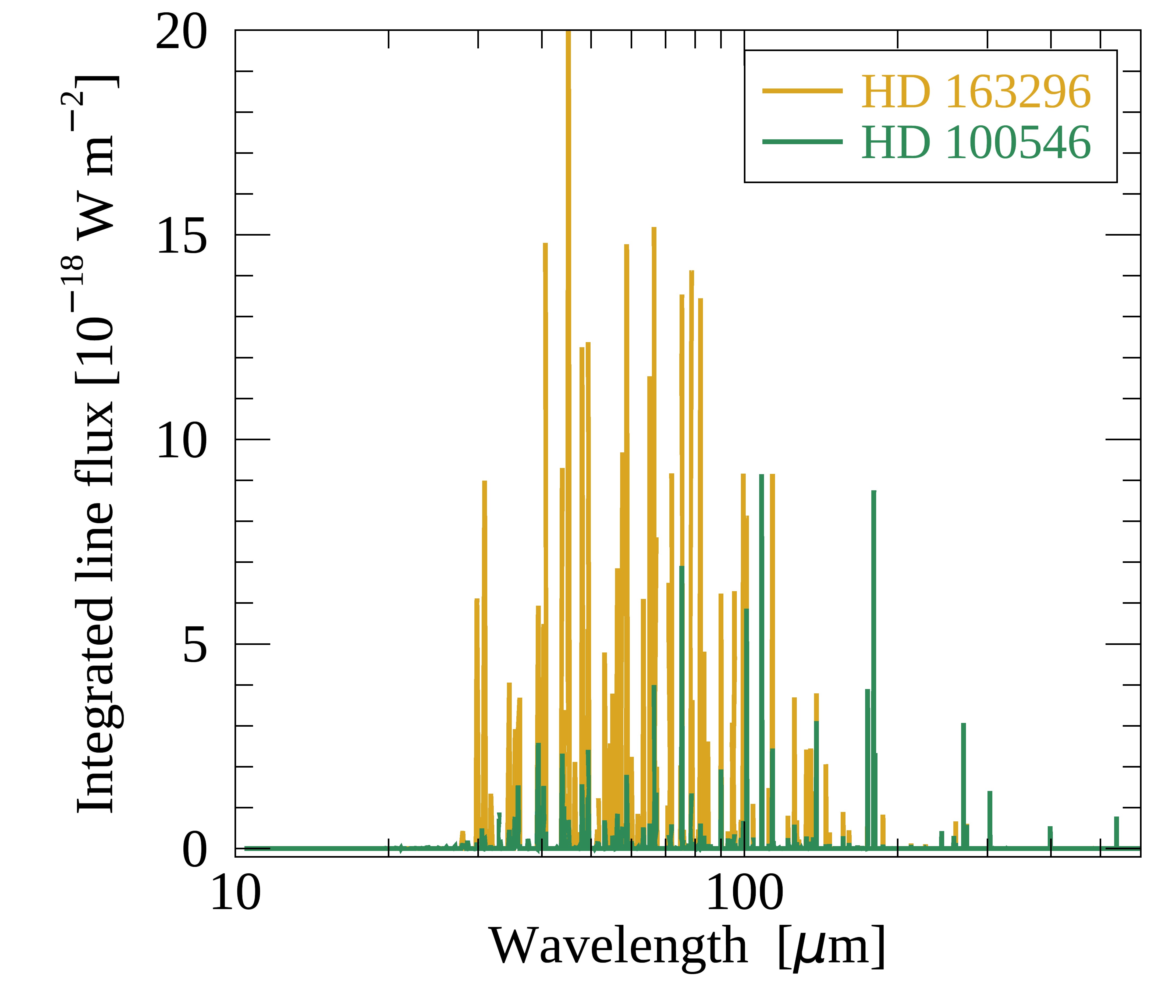}
    \caption{Synthetic \water \ spectrum of HD 163296 and HD 100546 based on the \water \ parametric distribution reported in Figure~\ref{fig:final}.}
    \label{fig:synthetic}
\end{figure}

Figure~\ref{fig:synthetic} shows the synthetic \water \ spectra for both discs based on the DALI parametric distribution reported in Figure~\ref{fig:final}. The spectroscopic data for the ray tracing are from the LAMDA database \citep[][and reference therein]{Schoier05}. Submillimetre observations with e.g. ALMA are potentially able to detect emission of \water \ isotopologues, such as H$_2^{18}$O. According to our predictions, the brightest \water \ submillimetre transition is the $4_{14}-3_{21}$ (lower energy level of 212\,K); while the main isotopologue cannot be observed from the ground because of atmospheric absorption, the corresponding H$_2^{18}$O transition at 390,607\,GHz falls in the ALMA band 8. Assuming a H$_2$O / H$_2^{18}$O abundance ratio of 560, the predicted flux of the H$_2^{18}$O $4_{14}-3_{21}$ transition is $7.5 \times 10^{-24}\,$W\,m$^{-2}$ and $1 \times 10^{-24}\,$W\,m$^{-2}$ for HD 163296 and HD 100546, respectively. These lines are too faint to be detected with current instrumentation in a reasonable amount of time. 

\end{appendix}

\end{document}